%% file: MFE-paper_TR.tex
\documentclass[opre,nonblindrev,copyedit]{informs2_TR}
\usepackage{graphics}
\usepackage{epsfig}
\usepackage{amsmath}
\usepackage{amssymb}
\usepackage{amsbsy}
\usepackage{setspace}

\input{OE-shortcuts}

\usepackage{natbib}
 \bibpunct[, ]{(}{)}{,}{a}{}{,}%
 \def\bibfont{\small}%
 \def\bibsep{\smallskipamount}%
 \def\bibhang{24pt}%

\TheoremsNumberedThrough     % Preferred (Theorem 1, Lemma 1, Theorem 2)
\ECRepeatTheorems
\EquationsNumberedThrough    % Default: (1), (2), ...

\begin{document}

\input{abstract}

\maketitle

%\begin{abstract}
%  We study stochastic dynamic games with a large number of players,
%  where players are coupled via their payoff functions.  We consider
%  {\em mean field equilibrium} for such games: in such an equilibrium,
%  each player reacts to only the long run average state of other players.
%  In this paper we focus on a special class of stochastic games, where
%  a player experiences strategic complementarities from other players;
%  formally the payoff of a player has increasing differences between
%  her own state and the aggregate empirical distribution of the states
%  of other players.  We find necessary conditions for the existence of
%  a mean field equilibrium in such games. Furthermore, as a simple consequence of this
%  existence theorem, we obtain several natural monotonicity
%  properties.  We show that there exist a ``largest''
%  and ``smallest'' equilibrium among all those where the equilibrium
%  strategy used by a player is nondecreasing.  We also show that
%  natural best response dynamics converge to each of these
%  equilibria.

\input{introduction}

\input{model}

\input{existence_cd}

\input{convergence}

\input{comparative_statics}

\input{actions}

\input{existence}

\input{examples}

\input{numerics}

\input{conclusion}

\input{appendix}

%%% \section{Conclusions}
%%% In this paper, we study stochastic games with a large number of players, where players are coupled via their payoff function. Similar to~\cite{weintraub_2008}, we showed that for a certain class of stochastic games, we can approximate MPE by a computationally
%%% simpler concept called oblivious equilibrium. We address two key questions pertinent to oblivious equilibrium : the existence of an oblivious equilibrium and its approximation in finite systems. We establish key properties that are required to ensure the existence of an oblivious equilibrium in a large class of stochastic games. We also show that under minor strengthening of these conditions, not only an oblivious equilibrium exists, it is also a good approximation to Markov perfect equilibrium. This paper thus develops a unified framework to study oblivious equilibrium and extends previous known results in this area.

%%%%%%%%%%%%%%%%%%%%%%%%%%%%%%%%%%%%%%%%%%%%%%%%%%%%%%%%%%%%%%%%%%%%%%%%%%%%%%%%%%%%%%%%%%%%%%%%%%%%%%%%%%%%%%%%%%%%%%%%%%%%%%%%%%
\bibliographystyle{ormsv080}
\bibliography{OE-paper}

\end{document}

%% file: OE-shortcuts.tex
\usepackage{epsfig}
\usepackage{graphicx}
\usepackage{makeidx}
\usepackage{multicol}
\usepackage[usenames]{color}
\usepackage{amsmath,amssymb}

\definecolor{ForestGreen}{RGB}{34,139, 34}

\newcommand{\newsa}[1]{#1}

\newcommand{\E}{\ensuremath{\mathbb{E}}}

\newcommand{\R}{\ensuremath{\mathbb{R}}}
\newcommand{\reals}{\ensuremath{\mathbb{R}}}
\newcommand{\prob}{\mathop{\mathrm{Prob}}}
\renewcommand{\P}{\ensuremath{\mathbb{P}}}

\newcommand{\SD}{\ensuremath{\mathsf{SD}}}

\renewcommand{\mc}[1]{\ensuremath{\mathcal{#1}}}
\newcommand{\mbf}[1]{\ensuremath{\mathbf{#1}}}

\newcommand{\mfr}[1]{\ensuremath{\mathfrak{#1}}}

\newcommand{\ul}[1]{\ensuremath{\underline{#1}}}
\newcommand{\ol}[1]{\ensuremath{\overline{#1}}}

\newcommand{\delnote}[1]{}
\newcommand{\newrj}[1]{ #1}

\newcommand{\fmi}{\ensuremath{\boldsymbol{f}^{(m)}_{-i, t}}}

\newcommand{\indic}[1]{\ensuremath{\textbf{1}_{#1}}}

\renewcommand{\v}[1]{\ensuremath{\boldsymbol{#1}}}

\mathchardef\mhyphen="2D

\newcommand{\ignore}[1]{}

%% file: abstract.tex
\RUNAUTHOR{Adlakha and Johari}
\RUNTITLE{Mean Field Equilibrium in Dynamic Games with Complementarities}

\TITLE{Mean Field Equilibrium in Dynamic Games with Complementarities}

\ARTICLEAUTHORS{%
\AUTHOR{Sachin Adlakha}
\AFF{Department of Electrical Engineering, Stanford University,
  Stanford, CA, 94305, \EMAIL{adlakha@stanford.edu}} %, \URL{}}
\AUTHOR{Ramesh Johari}
\AFF{Department of Management Science and Engineering, Stanford
  University, Stanford, CA, 94305, \EMAIL{ramesh.johari@stanford.edu}}
% Enter all authors
} % end of the block

\ABSTRACT{ We study a class of stochastic dynamic games that exhibit
  {\em strategic complementarities} between players; formally, in the
  games we consider, the payoff of a player has increasing differences
  between her own state and the empirical distribution of the states
  of other players. \newrj{Such games can be used to model a diverse set of
  applications, including network security models, recommender
  systems, and dynamic search in markets.   Stochastic games are generally
  difficult to analyze, and these difficulties are only exacerbated when
  the number of players is large (as might be the case
  in the preceding examples).}

%In several such games, a player experiences strategic complementarities from other players;
%For example, in a model of interdependent security games~\cite{kunreuther_2003}, a player's  return on its own investment in security is increased if many other players invest in their own security.
We consider an approximation methodology called {\em mean field
  equilibrium} to study these games. In such an equilibrium, each
player reacts to only the long run average state of other
players.  We find necessary
conditions for the existence of a mean field equilibrium in such
games. Furthermore, as a simple consequence of this existence theorem,
we obtain several natural monotonicity properties.  We show that there
exist a ``largest'' and a ``smallest'' equilibrium among all those where
the equilibrium strategy used by a player is nondecreasing, and we also
show that players converge to each of these equilibria via
natural \newrj{{\em myopic learning dynamics}; as we argue, these dynamics
are more reasonable than the standard best response dynamics.} We also provide sensitivity
results, where we quantify how the equilibria of such games move in
response to changes in parameters of the game (e.g., the introduction of
incentives to players).
}

%% file: introduction.tex
\section{Introduction}   
\label{sec:intro}

%Systems with large number of interacting agents are common in a 
%variety of engineering, social, industrial, and economic settings. 
%In a diverse set of applications such as power control in wireless 
%networks, or competition among firms in a market, there are large number 
%of agents interacting with each other. In many of these applications, these
%agents interact in a stochastic dynamic environment. A natural modeling framework
%to study such interactions is that of {\em stochastic games}.

This paper studies a class of games that exhibit {\em strategic
  complementarities} between players.  A strategic complementarity
exists if, informally, ``higher'' actions by other players increase
the return to higher actions for a given player.  Games
with strategic complementarities are a powerful modeling tool,
applicable in a wide range of situations, including: systems with positive network effects (such as network
security models, recommender
systems, and social networks); coordination problems; dynamic search
in markets; social learning; and oligopoly models (e.g., quantity or
price competition with complementarities).

Our focus in this paper is on {\em dynamic} games with strategic
complementarities.  Strategic complementarities have long provided a fertile analytical
ground for static game theoretic models; see, e.g., \cite{Milgrom90,
  Vives90}, and \cite{topkis_1998}.  However, the literature on
dynamic games with
complementarities has emerged relatively recently by comparison. Much
of the attention in prior work on such games  
has focused on developing existence proofs for equilibrium; see, e.g.,
\citet{Curtat96, Amir02, Amir_2005, sleet_2001,
  vives_2009} for these results.

\newrj{In this paper we consider a class of dynamic games referred to
  as {\em stochastic games}; in these games agents' actions directly
  affect underlying state variables that influence their payoff
  \citep{shapley_1953}.}  The standard solution concept for stochastic
games is {\em Markov perfect 
  equilibrium} \citep{fudenberg_1991}.  Despite the previously cited existence results
for Markov perfect equilibria in games with complementarities, there remain two significant
obstacles, particularly as the number
of players grows large.  First is {\em computability}: the state space
of the preceding games expands in dimension with the number of
players, and thus the ``curse of dimensionality'' kicks in, making 
computation of Markov perfect equilibria essentially infeasible \citep{pakes_2001, pakes_2006}.  Second is {\em
  plausibility}: as the number of players grows large, it becomes
increasingly difficult to believe that individual players track the
exact behavior of the other agents.
Rather than treat the growth of
the population as an impediment to analysis, this paper addresses
these obstacles by exploiting an asymptotic regime where the number of
players grows large to {\em simplify} analysis of equilibria.

We consider an approximation methodology where agents optimize only
with respect to long run average estimates of the distribution of
other players' states, that we refer to as {\em mean field
  equilibrium}; this notion has been utilized across a range of work
in economics, operations research, and control (as we discuss below).
In a mean field equilibrium, individuals take a simpler view of the world: they
postulate that fluctuations in the empirical 
distribution of other players' states have ``averaged out'' due to
large scale, and thus optimize holding the state distribution of other
players fixed.  Mean field equilibrium requires a consistency check:
the postulated state distribution must arise from the optimal
strategies agents compute.

\newrj{Our results provide valuable insight into the structure of mean
field equilibria in such games, as well as computational tools to
determine such equilibria.
To motivate our results, we first provide
several examples of stochastic games with complementarities where the
approach taken in this paper applies.  These examples---particularly
the first four---often exhibit large numbers of players, and thus the
benefits of mean field equilibrium are significant.  We demonstrate in Section
\ref{sec:examples} that each of these examples can be analyzed using
the results we develop in this paper.}

\begin{example}[Interdependent security]  
\label{ex:ids}
In {\em interdependent security} games, as introduced
in \cite{kunreuther_2003}, a large number of agents
make individual decisions about their own security.  However, the
ultimate security of an agent depends on the security decisions made
by other agents. For example, imagine a network of computers where
each individual user makes an investment in keeping her own machine
secure. This investment may be in the form of advanced anti-virus
filters, firewalls, etc.  While these investments improve the security
of the individual computer, it can still be affected if the other
computers in the network are not properly secured.  In the
interdependent security games we consider, agents take actions at some
cost to improve their own security level, and earn a payoff each
period that depends on whether or not a security breach occurs.  The
fact that the probability of a security breach is influenced by
others' security levels introduces strategic complementarities into
the stochastic game.
\hfill \halmos \end{example}

\begin{example}[Collaborative filtering]
\label{ex:cf}
Many large online recommendation systems, such as those used by Netflix
and Amazon, rely on collaborative
filtering.  In such
systems, if an individual puts forth greater effort in maintaining
their profile, the recommendations {\em they} receive will improve.
  However, the recommendations {\em other} individuals receive improve
  as well, and typically other individuals will feel a stronger
  incentive to exert additional effort to maintain their profile in
this case.  In the absence of such effort, the profile of an agent
becomes stale and useless both to her and others in the system.  Thus
collaborative filtering systems exhibit strong strategic
complementarities.
\hfill \halmos \end{example}

\begin{example}[Dynamic search with learning]
\label{ex:dsl}
In dynamic search models, traders in a market exert effort to find
trading partners \citep{diamond_1982}.  Such models are commonly used
to study, e.g., decentralized matching in labor markets. We consider a model 
where at each time step, traders also gain experience by exerting
effort; this experience makes future effort more productive.
Of course traders' experience increases
as {\em they} put forth more effort; but their experience also increases as 
{\em others} put forth more effort since this increases the likelihood of useful
interactions per unit effort. This creates strategic complementarities
between the players; such a model was considered by 
\cite{Curtat96}.
\hfill \halmos \end{example}

\begin{example}[Coordination games]
\label{ex:coord}
There exist many examples in operations and economics where agents are
trying to {\em coordinate} on a common goal; for example, this is the
case when firms try to coordinate on a common standard.  In a
coordination game, a collection of agents take individual actions to
converge on a common state. One such stylized model is the linear-quadratic
decentralized coordination problem studied by \cite{huang_2005}.
Agents can change their state by exerting effort at some cost.
\newrj{Further, each agent incurs an additional state-dependent cost each
time period; this cost is quadratic in the distance to the {\em
  average} of other agents' states.}  This type of game can be shown to exhibit strategic complementarities
between agents.
\hfill \halmos \end{example}

\begin{example}[Oligopolies and complementary goods]
\label{ex:cg}
Consider competition among firms producing complementary goods. In
particular, suppose firms have effective monopolies in their own
markets, but that their goods are complements, so that the consumption
of one good will increase the demand and consumption of others.  Such
models naturally exhibit strategic complementarities.

One potential issue in using mean field models to analyze oligopolies
with complementary goods is that the number of firms may not be too
large, thus raising questions about the validity of a mean field limit
in the first place.  However, even in such a setting mean field models
have value, because they provide structural insight into optimal
strategies under a model of rationality that is perhaps more
plausible, as discussed above. Indeed, econometric
analysis using mean field models of dynamic oligopolies has proven
valuable for a range of industries with relatively small numbers of
firms (see \cite{weintraub_2010} for examples).
\hfill \halmos \end{example}

%
%The standard solution concept for stochastic games is {\em Markov perfect
%  equilibrium} (MPE) \citep{fudenberg_1991}.  Despite existence results
%for MPE in games with complementarities, there remain two significant
%obstacles, particularly as the number
%of players grows large.  First is {\em computability}: the state space
%of the preceding games expands in dimension with the number of
%players, and thus the ``curse of dimensionality'' kicks in, making MPE
%computation essentially infeasible \citep{pakes_2001, pakes_2006}.  Second is {\em
%  plausibility}: as the number of players grows large, it becomes
%increasingly difficult to believe that individual players track the
%exact behavior of the other agents.  Rather than treat the growth of
%the population as an impediment to analysis, this paper addresses
%these obstacles by exploiting an asymptotic regime where the number of
%players grows large to {\em simplify} analysis of equilibria.

% there remain significant obstacles to computation of and convergence
% to equilibria, as well as characterization of sensitivity of 
% equilibria to parameters of the game.  These problems are particularly
% acute in games with a large number of players.
%which exhibit
%two additional salient features.   The second salient feature of our model is that we consider
%Our goal is to simplify the analysis of stochastic games with
%complementarities, such as those described in the preceding examples.

\newrj{Our main results provide conditions that ensure existence of mean
field equilibria in stochastic games with complementarities.  We also
establish that simple learning procedures converge to equilibria, and
provide insight into sensitivity of equilibria to parameter changes.
\newrj{We consider a general class of models with parsimonious
  assumptions over model primitives that ensure strategic
  complementarities.  In particular, our model class allows players to
  be coupled {\em both} via their payoff function and state
  transitions, i.e., players' payoffs and state transitions can depend
  on states or actions of other players.  We also discuss
  extensions of our
  results to models with multidimensional state and action spaces, and
  with heterogeneity among players.}  Details of our results follow.}

% % \delsa{In this paper we study mean field equilibrium of stochastic games with complementarities. 
% % Inspired by the preceding set of examples, we consider a class of
% % games where the players have individual state and they take actions
% % to change their state at each time period. Furthermore,
% % the players are coupled via their payoff function and transition kernel; a player's payoff
% % depends on the states of other players, and a player's state
% % transitions may be influenced by the states of other players.  We develop several
% % parsimonious assumptions over the state dynamics and payoff function
% % that ensure strategic complementarities among 
% % the players. Notably, these assumptions are satisfied for natural
% % instances of the examples mentioned above.}

% In this paper, we study mean field equilibrium for a class of games that exhibit strategic complementarities.
% Compared to existing literature, we provide existence and convergence results for a much broader class of stochastic games. Specifically, we consider a class of games where the players have individual state and they take actions
% to change their state at each time period. We develop several
% parsimonious assumptions over the state dynamics and payoff function
% that ensure strategic complementarities among 
% the players. Notably, these assumptions are satisfied for natural
% instances of the examples mentioned above and they also generalize existing works on mean field equilibrium for stochastic games.}

\begin{enumerate}
\item {\em Structural characterization of mean field equilibrium.}  We establish
  existence of a mean field equilibrium in a general stochastic game model using lattice
  theoretic techniques.  Lattice theoretic methods are typically
  applied in games with complementarities; the key techniques we use
  are due to \cite{tarski_1955, kamae_1977, Hopenhayn92, Zhou94}, and
  \cite{topkis_1998}.  Despite the use of lattice theoretic techniques
  in our analysis, existence of equilibria in our game cannot be inferred from
  existence results for other games in the literature.  Moreover, we
  show that there exists a ``largest'' and ``smallest'' equilibrium
  among the set of all mean field equilibria with nondecreasing strategies.  Thus, in
  particular, there is a natural dominance relationship among the mean field equilibria
  of a given stochastic game with complementarities.  This is
  particularly valuable in dynamic games, because our characterization
  applies to the {\em distribution} of states of agents in
  equilibrium.  

\newrj{We note that prior literature has established existence of
  equilibrium in stochastic games with complementarities; however, these results
  typically also require use of topological fixed-point theorems such
  as Kakutani's theorem \citep{Curtat96, Amir02, Amir_2005}.  More
  closely related to our paper is the work of \cite{sleet_2001}, who
  considers mean field equilibria of a dynamic price-setting game
  with stochastic, exogenous firm-specific demand shocks per period.
  The general analytical techniques in this paper can be applied to recover
  the existence result for that game.}

\item {\em Convergence to equilibrium.}  We provide two convergence
  results.  First, we study a standard {\em best response dynamic} (BRD).
  In this algorithm, at each time step, each agent computes the
  stationary population state distribution that would be induced by
  the current strategies of others, and in turn computes the best
  response to that state distribution.  Using monotonicity properties
  derived in establishing the existence of mean field equilibrium, we show that BRD
  converges.

However, BRD is unsatisfying both computationally and practically.
From a computational standpoint, BRD requires computation of a
stationary distribution given the current strategy choices of agents
in the system; this is in principle a complex procedure to execute at
each iteration.  More importantly, BRD is an implausible approach to
play in an actual game: it is unlikely that agents would explicitly
compute the stationary distribution their competitors would obtain.

Instead, we consider a more a natural form of {\em myopic learning
  dynamics} (MLD) among
the players; convergence of MLD is a central insight of our
paper.  In particular, suppose that initially, each agent starts at
the lowest (resp., highest) possible state.  At each time step, agents
observe the current empirical population state distribution, and
conjecture that this distribution will {\em remain constant for all
  time}; with this conjecture they compute an optimal strategy, and
play in the next period according to that strategy.  At the next time
step, the state distribution will evolve, and agents repeat the same
heuristic.  We show that this dynamic converges to the lowest (resp.,
highest) mean field equilibrium among all equilibria with nondecreasing strategies.

Note that MLD resolves both the {\em computability} and {\em
  plausibility} issues raised above.  First, it is a natural, simple,
implementable algorithm for finding a mean field equilibrium; indeed, MLD has some
similarities with model predictive control or receding horizon control
\citep{garcia_1989}, both popular approaches to complex dynamic control
problems.  Second, it corresponds to a learning dynamic that demands
only a weak form of rationality and forecasting from the players, and
yet yields an equilibrium in the limit.

\item {\em Separable stochastic games.}  Although appealing,
  the general theory does pose some significant issues in application:
  the complementarity requirements on model primitives may preclude
  important and interesting cases of practical interest.
  Complementarity is a strong requirement, but also 
  brittle: a model that does not appear to satisfy the assumptions
  {\em a priori} may do so through a judicious change of variables. 
\newrj{We employ this fact to show that a range of games that do not
satisfy the assumptions of our baseline model can be studied by a
suitable change of variables, provided that the payoff
is separable in the state and action of a given player---often a
relatively mild assumption. Notably, models with {\em linear} dynamics fall
in this class.  This greatly expands the 
set of models that can be analyzed within our framework.}

\item {\em Sensitivity.}  Finally, essentially for free, the
  complementarity structure allows us to analyze changes in the
  equilibrium in response to changes in parameters of the game.  In
  particular, we can predict shifts (in a first order stochastic
  dominance sense) of the equilibrium state distribution of players in
  response to exogenous parameter changes.  Such
  sensitivity analysis, or {\em comparative statics}, allows our model
  to address, e.g., the value of incentives to increase security
  levels, or the value of increasing the quality of recommendations by
  a given factor.
\end{enumerate}

The remainder of the paper is organized as follows.  In Section
\ref{sec:model} we introduce our basic stochastic game model as well
as the formal definition of mean field equilibrium.  Notably, we also
discuss a justification for the use of mean field equilibrium: that it approximates
equilibria of finite games well.  This approximation property has been
developed in a variety of specific contexts in the past (see, e.g.,
\citealp{glynn_2004}, \citealp{huang_2006}, \citealp{weintraub_2008}, and
\citealp{tembine_2009}), and in our context we apply the methodology
developed in \cite{adlakha_2010} (inspired by \citealp{weintraub_2008})
to justify mean field equilibrium as a limiting notion of equilibrium.

Next, in Section \ref{sec:existence-cd}, we define stochastic games
with complementarities.  We then prove our first main result: that a mean field equilibrium exists for a stochastic game with complementarities.  In Section
\ref{ssec:ordering-cd}, we show that equilibria are ``ordered,'' in
the sense that there exists a smallest and largest mean field equilibrium among all those
where the equilibrium strategy is nondecreasing.  In Section
\ref{sec:convergence}, we prove convergence of both the BRD and MLD
algorithms described above.  We also discuss the performance of MLD in
finite systems.  

In Section~\ref{sec:comp-stat}, we provide
comparative statics results for the games under consideration.  In
Section~\ref{sec:actions}, we extend our results to cover games where
players' payoffs and transition kernels may depend on the {\em
  actions} of others, rather than their states.  In Section~\ref{sec:existence} we consider separable stochastic games with
complementarities (as described above), and establish that these are a
special case of our basic model of stochastic games with
complementarities.

In Section~\ref{sec:examples}, we revisit each of the examples
described above.  In particular, we provide formal verification that
these examples satisfy the assumptions made in the paper to obtain
existence and convergence results.  Finally, in Section~\ref{sec:numerical}, we study a particular instance of an
interdependent security game.  We use this game to illustrate several
computational insights, including verification of comparative statics
results, as well as exploration of the performance of the MLD dynamic
described above.  Section~\ref{sec:conclusion} concludes with a
discussion of extensions to include both player heterogeneity (i.e.,
type information) and multidimensional state and/or action spaces.

We conclude by surveying related work on mean field equilibrium.  The
notion of mean field equilibrium is inspired by mean field models in
physics, where large systems exhibit macroscopic behavior that is
considerably more tractable than their microscopic description.  (See,
e.g., \cite{mezard_2009} for background, and \cite{blume_1993} and
\cite{morris_2000} for related ideas applied to static games.)  In the 
context of stochastic games, mean field equilibrium and related approaches have been
proposed under a variety of monikers across economics and engineering;
see, e.g., studies of anonymous sequential games
\citep{jovanovic_1988, bergin_1995}; stationary equilibrium
\citep{hopenhayn_1992}; dynamic stochastic general equilibrium in
macroeconomic modeling \citep{stokey_1989}; Nash certainty equivalent
control \citep{huang_2006, huang_2007}; mean field games
\citep{lasry_2007}; oblivious equilibrium \citep{weintraub_2008,
  weintraub_2010}; and dynamic user equilibrium \citep{friesz_1993, wunderlich_2000}.  Mean field equilibrium has also been studied in
recent works on information percolation models \citep{duffie_2009},
sensitivity analysis in aggregate games \citep{acemoglu_2009},
coupling of oscillators \citep{yin_2010}, and in
scaling behavior of markets \citep{bodoh_creed_2010}.

% In systems involving large number of interacting players, one is often interested in understanding how the equilibrium behaves when certain incentives are introduced in the payoff function. For example, in the security games, one would like to understand the following question: {\em How does the investment strategy of an agent and the global security profile of the entire population change, when individual players are incentivized to invest in security?} Or in the context of coordination games one would like to answer the following question: {\em How does the optimal trajectory of the population mass change when the marginal cost of action is reduced?} We provide answers to such questions via our results on comparative statics for games with complementarities. 

%The remainder of the paper is organized as follows.  In Section
%\ref{sec:model}, we introduce our stochastic game model.  We also
%define and discuss the notion of mean field equilibrium.  In Section
%\ref{sec:existence}, we prove that a mean field equilibrium exists for
%our stochastic game, and characterize the largest and smallest
%equilibria.  In Section
%\ref{sec:convergence}, we prove natural best response dynamics converge to these
%equilibria. Section~\ref{sec:comp-stat} provides results on comparative statics for such games. Finally, in Section \ref{sec:extensions} we conclude with
%several extensions.  In particular, we discuss a theorem that shows MFE is a good
%approximation to MPE when the number of agents in the system grows
%large \cite{adlakha_2010}.  We also discuss generalization of our
%baseline model.  

%% file: model.tex
\section{Model and Definitions}
\label{sec:model}

In this section we begin with preliminaries.  We define a general 
model of a stochastic game in Section \ref{subsec:stochastic-game}; in
the games we consider, agents take actions to update their own states,
and their payoffs and state transitions may be affected by the states
of others. Next, in Section \ref{subsec:OE}, we define {\em mean field
  equilibrium}, \newrj{and in Section \ref{subsec:approx} we provide a formal
justification for mean field equilibrium as an approximation to
equilibria in games with a large finite number of players.} Finally, in Section \ref{ssec:prelims}, we discuss
lattice-theoretic preliminaries necessary for the analysis in the
sequel.  

\subsection{Stochastic Games}

\label{subsec:stochastic-game}
We consider a
game played among $m$ players.
A {\em stochastic game} is a tuple $\Gamma = (\mc{X}, \mc{A}, \P,
\pi, \beta)$ defined as follows.

{\em Time.} The game is played in discrete time, with time
periods by~$t = 0, 1, 2, \ldots$.%\in \Z_{+}$.

{\em State.} The state of player $i$ at time $t$ is denoted by
$x_{i,t} \in \mc{X}$, where $\mc{X} \subset \reals$ is 
  compact.   We use~$\v{x}_{-i,
  t}$ to denote the state of all players except player~$i$ at
time~$t$.

\textit{Action.} The action taken by player $i$ at time $t$ is denoted
by $a_{i,t}$.  The set of feasible actions when the player is in state
$x$ is a compact set $\mc{A}(x) \subset \reals$.  We
let $\mathsf{A} = \cup_{x \in \mc{X}} \mc{A}(x)$, and assume that
$\mathsf{A}$ is compact as well.

\textit{Transition probabilities.} The state of a player evolves
according to the following Markov process. If the state of player $i$
at time $t$ is $x_{i,t} = x$, the player takes an action $a_{i,t} =
a \in \mc{A}(x)$ at time $t$, and the state of every other player at time~$t$ is $\v{x}_{-i,t} = \v{y}$, then the next state is distributed according to the
Borel probability measure $\P(\cdot|x, a, \v{y})$, where for Borel sets $S \subset
\mc{X}$,
\begin{align}
\label{eqn:tm}
\P(S| x, a, \v{y}) = \prob\left(x_{i,t+1} \in S |  x_{i,t} = x, a_{i,t} = a, \v{x}_{-i,t} = \v{y}\right).
\end{align}
Further, given $x_{i,t}$, $a_{i,t}$, and $\v{x}_{-i,t}$, the next
state $x_{i,t+1}$ is conditionally independent of all other past
history of the game.

\textit{Payoff.} The single period payoff to player~$i$ at time~$t$ is
$\pi(x_{i,t},a_{i,t}, \v{x}_{-i,t}) \in \R$.  Note that all players have the same
payoff, and it is independent of the actions taken by other players. 

\textit{Discount factor.} The players discount their future payoff by
a discount factor $0 < \beta < 1$.  Thus a player $i$'s infinite
horizon payoff is given by:
\[ \sum_{t = 0}^\infty \beta^{t}\pi(x_{i,t}, a_{i,t}, \v{x}_{-i,t}). \]
\vspace{0.1in}
%\item \textit{Initial State.} We assume that the initial state of a player is chosen according to the probability mass function $\P_{0}$, where $\P_{0}(z)= \prob\big(x_{i,0} = z\big)$.

It may initially appear unusual that we do not include the number of
players as part of the specification of the game; however, this choice
is deliberate.  We ultimately study $\Gamma$ in a limiting regime where the
number of players grows large, and as a result, mean field
equilibrium is defined without regard to a fixed finite number of
players.  For this reason we do not include $m$ in the tuple defining
$\Gamma$.  (See the next section for further discussion of the
motivation for mean field equilibrium.)

In the model described above, the players are coupled to each other
via their state evolution and the payoff function. In a variety of games, this coupling between
players is independent of the identity of the players. The notion of
\textit{anonymity} captures scenarios where the interaction between
players is via aggregate information about the state. Let $\fmi(y)$
denote the fraction of players (excluding player $i$) that have their
state as $y$ at time $t$, i.e.:
\begin{align}
\label{eqn:actual-dist}
\fmi(y) = \frac{1}{m-1}\sum_{j \neq i}\indic{\{x_{j,t} = y\}},
\end{align}
where $\indic{\{x_{j,t} = y\}}$ is the indicator function that the
state of player $j$ at time $t$ is $y$.  We refer to $\fmi$ as the
{\em population state} at time $t$ (from player $i$'s point of view).

\begin{definition}[Anonymous Stochastic Game]
\label{def:mean-field-games}
A stochastic game $\Gamma = (\mc{X}, \mc{A}, \P,
\pi, \beta)$ is called an \textit{anonymous stochastic game} if
the transition probability measure and payoff for player~$i$ depend on
$\v{x}_{-i,t}$ only through $f_{-i,t}$.  Through an abuse of notation,
we write the transition probability measure as $\P(\cdot |
x_{i,t}, a_{i,t}, \fmi)$ and the payoff function of player $i$ as
 $\pi(x_{i,t},a_{i,t}, \fmi)$.
\end{definition}

\newsa{The examples discussed in the Introduction naturally belong to the class of
anonymous stochastic games. For example, in the interdependent
security model (Example \ref{ex:ids}), it is natural to assume that a
single player's payoff is affected by the empirical distribution of
security levels of other players in the network, but not by their specific identity.
The same assumption is also plausible for the other examples presented
earlier.}

For the remainder of the paper, we focus our attention on anonymous
stochastic games. For ease of notation, we often drop the subscript
$i$ and $t$ to denote a generic transition probability measure and a generic payoff, i.e., we denote a generic transition probability measure by $\P(\cdot | x, a, f)$ and a generic payoff by $\pi(x, a, f)$, where $f$ represents the
population state of players other than the player under
consideration.  We let $\mfr{F}$ denote the 
set of all Borel probability measures on $\mc{X}$.

\subsection{Mean Field Equilibrium}
\label{subsec:OE}

In a game with a large number of players, we might expect that
fluctuations of players' states ``average out'', and hence the actual
population state remains roughly constant over time.  Because the
effect of other players on a single player's payoff is only via the
population state, it is intuitive that, as the number of players
increases, a single player has negligible effect on the outcome of the
game. This intuition is formalized through the notion of {\em mean
  field equilibrium} \citep{jovanovic_1988, bergin_1995,
  hopenhayn_1992, stokey_1989, friesz_1993, huang_2006, huang_2007,
  lasry_2007, weintraub_2008,  weintraub_2010, adlakha_2010, bodoh_creed_2010}.

In mean field equilibrium, each player optimizes its payoff based on
only the long-run average population state. Thus, rather than keep
track of the exact population state, a single player's action depends
only on her own current state as well as the long run average
population state.  This is motivated by the fact that a single player
need not concern herself with the fine scale dynamics of competitors' specific
states.  Given this simplified player behavior, note that each player
must solve a dynamic program to determine their optimal strategy; the
strategy chosen by each player then leads to a long-run average
population state.  Mean field equilibrium requires that the latter
long-run average population state matches the original conjecture made
by the players.

Note that in a mean field equilibrium, because players optimize holding the population
state constant, their optimal strategies will depend
only on their current state.  We call such players {\em
  oblivious}, and refer to their strategies as {\em oblivious strategies}.
%Such policies are referred to as
%\textit{oblivious policies}. The term ``oblivious'' is used to signify
%the fact that the policy does not take into account the complete state
%of the competitors.
This approach does not require players to be aware of each others'
exact states, if every player is aware of the long-run average population
state. Furthermore, observe that if all players are oblivious,
players' states evolve independently.

In this section we fix an anonymous stochastic game $\Gamma = (\mc{X}, \mc{A}, \P,
\pi, \beta)$.  Formally, an {\em oblivious strategy} is a strategy that
depends only on the player's current state.  We let $\mfr{M}_O$ denote
the set of oblivious strategies.

\begin{definition}
Let $\mfr{M}_O$ be the set of oblivious strategies available to a
player:
\begin{align}
\label{eqn:policy-set-OE}
\mfr{M}_O = \left\{ \mu : \mc{X} \to \mathsf{A}\ |\ \mu(x) \in \mc{A}(x)\
\text{for all}\ x \in \mc{X}\right\}.
\end{align}
\end{definition}

%Consider a player and denote by~$f$ the long-run average
%population state.
Given an oblivious strategy $\mu \in \mfr{M}_O$, a
player~$i$ takes an action~$a_{i,t} =
\mu(x_{i,t})$ at time~$t$.  If the player conjectures the aggregate
population state to be $f$, then she also conjectures that her next
state is randomly distributed according to the transition probability measure~$\P$:
\begin{equation}
\label{eq:oe_dynamics}
x_{i,t+1} \sim \P( \cdot | x_{i,t}, \mu(x_{i,t}), f),
\end{equation}
where $f$ is the conjectured long run average population state. 

We define the
\textit{oblivious value function}~$V(x|\mu,
f)$ to be the expected net present value for any player with
initial state~$x$, when the long run average population state
is conjectured to be $f$, and the player uses an oblivious strategy~$\mu$. We have
\begin{align}
\label{eqn:oe-value-func}
V\left(x | \mu, f\right) \triangleq 
\E\left[\sum_{t=0}^{\infty}\beta^{t}\pi(x_{t}, a_t, f) \ \Big| \ x_{0} = x;\ \mu \right]. 
\end{align}

%\addnote{Should we change this to be arbitrary f and not $\tilde{f}$?}

Given a population state $f$, a player computes an
optimal strategy by maximizing their oblivious value function.  Note
that because the oblivious value function does not track the evolution of the
population state, we should expect a player's optimal
strategy to depend only on their current 
state---i.e., it must be oblivious.  We capture this optimization
step via the operator $\mc{P}$ defined next.  

Define $V^*(x | f)$ as:
\[ V^*(x | f) = \sup_{\mu'\in
    \mfr{M}_O} V\left(x | \mu', f\right).\]

\begin{definition}
\label{def:P}
The operator $\mc{P}: \mfr{F}\rightarrow \mfr{M}_O$ maps a distribution
$f \in \mfr{F}$ to the set of optimal oblivious strategies. That
is, $\mu \in \mc{P}(f)$ if and only if
\begin{align*}
V\left(x | \mu, f\right) = V^*(x | f), \ \forall x
\in \mc{X}.
\end{align*}
\end{definition}
Note that in principle, $\mc{P}(f)$ may be empty, though
we show that under our assumptions this does not
occur.
%Further note that in general $\mc{P}(f)$ may be multiple
%valued; however, for the remainder of the paper, we assume that the
%$\mc{P}(f)$ is a singleton for every $f$ where it is
%nonempty.
%In
%Section~\ref{sec:extensions}, we extend our results to the case the
%policy is not unique.

%%% We often write $\mc{P}(f)$ to mean an
%%% optimal policy that maximizes the oblivious value function as given in
%%% equation~\eqref{eqn:oe-value-func} under the payoff $\pi(x, a,
%%% f)$. Sometimes we use the notation $\mc{P}_\pi(f))$ to
%%% make the dependence on the payoff explicit.

%A sin the case of MPE, we focus our attention on symmetric OE where each player uses the same policy~$\mu$. Note that under an oblivious policy, the state evolution of every player is independent. 

Now suppose that {\em all} players use the oblivious strategy
$\mu$, and the long run average population state~$f$ drives their
state dynamics.  In this scenario, we expect the long run 
population state to be an invariant distribution of the strategy $\mu$ under the dynamics
\begin{align}
\label{eq:oe_dynamics1}
x_{t+1}  \sim \P\left(\cdot | x_{t}, \mu(x_{t}), f\right).
\end{align}
We capture this relationship via the operator $\mc{D}$, defined next.

\begin{definition}
The operator $\mc{D} : \mfr{M}_O \times \mfr{F} \rightarrow \mfr{F}$ maps an
oblivious strategy $\mu$ and a distribution $f$ to the set of invariant distributions
associated with the dynamics~\eqref{eq:oe_dynamics1}.
\end{definition}
Thus, $g \in \mc{D}(\mu, f)$ if and only for all Borel sets $S \subset \mc{X}$,
\begin{align}
\label{eqn:inv-dist}
g(S) = \int_{\mc{X}}\P\left(S | x, \mu(x), f\right)g(dx)
\end{align}
Note that the image of the operator~$\mc{D}$ is empty if the strategy
does not result in an invariant distribution, though again, we show
under our assumptions that this does not occur.

We can now define mean field equilibrium.  If every agent
conjectures that $f$ is the long run population state, then
every agent would prefer to play an optimal oblivious strategy $\mu$.
On the other hand, if every agent plays $\mu$, and the long run
population state is indeed $f$, 
then $f$ must also be an invariant distribution of
\eqref{eq:oe_dynamics1}.  Thus mean field equilibrium requires a consistency
condition: the invariant distribution under $\mu$ and $f$ should be exactly
$f$.  
%%% Let us denote by~$q$ the
%%% steady state invariant distribution. For a player~$i$, the oblivious
%%% distribution~$f$ is the long-run average aggregate state
%%% distribution of its competitors and is given by
%%% \begin{align}
%%% \label{eqn:f-tilde}
%%% f(y) & \triangleq \E\big[\fmi(y)\ \vert \ \mu\big] \nonumber \\
%%% & \stackrel{(a)}= \E\Big[\frac{1}{m-1}\sum_{j \neq i}\indicf,\Big] \nonumber \\
%%% & \stackrel{(b)}= \frac{1}{m-1}\sum_{j \neq i}\prob(x_{j,t} = y) \nonumber \\
%%% & \stackrel{(c)}= q(y).
%%% \end{align}
%%% Here, the equality~(a) follows from the definition of~$\fmi$ as given in equation~\eqref{eqn:actual-dist} and the equality~(b) follows from the definition of the indicator function. The equality~(c) follows from the fact that if every player plays an oblivious policy~$\mu$, the states evolve independently and reach a steady state distribution~$q$ which is independent of the player.

\begin{definition}[Mean Field Equilibrium]
\label{def:oe}
A strategy~$\mu$ and a distribution~$f$ constitute a
mean field equilibrium if $\mu \in \mc{P}(f)$ and $f \in
\mc{D}(\mu, f)$.
\end{definition} 

%%% Thus, OE is a pair of a distribution and a policy vector that are
%%% consistent with each other. In other words, given the
%%% distribution~$f$, the policy~$\mum$ maximizes the oblivious
%%% value function as given in equation~\eqref{eqn:oe-value-func}, and the
%%% oblivious distribution~$f$ is the invariant distribution
%%% associated with the policy~$\mu$ when all players are
%%% oblivious and conjecture the state distribution to be $f$.

\subsection{The Approximate Markov Equilibrium Property}
\label{subsec:approx}

A natural question that arises in the context of mean field equilibrium is 
whether it is a good approximation to a game with finitely many
players. Here we present a  
formal justification for the notion of mean field equilibrium by
considering explicitly a limiting regime where the number of players
grows large.

Recall that we defined $\Gamma$
initially as a stochastic game with $m$ players.  The standard
solution concept for stochastic games is {\em Markov perfect
  equilibrium}.  In a Markov perfect equilibrium, players' strategies depend on their
own current state, as well as the current states of others; we refer
to such strategies as {\em cognizant} strategies.  This
larger state space makes Markov perfect equilibrium a much more complex equilibrium concept:
Markov perfect equilibrium is typically quite challenging to compute, and demands far greater
rationality on the part of the players.  

It can be shown, however, that under appropriate assumptions, a mean
field equilibrium is {\em approximately} a Markov perfect equilibrium as the number of
players grows large.  Formally, let $(\mu, f)$ be a mean field
equilibrium, and fix a single player $i$.  Suppose that we consider a
sequence of games with $m \to \infty$, where all players other
than player $i$ use the oblivious strategy $\mu$; and where the
initial state of all players other than player $i$ is sampled
i.i.d.~from $f$.  Then we can show
that as $m \to \infty$, the difference between the payoff player $i$
achieves by playing $\mu$ and the maximum possible payoff player $i$
can achieve by playing any cognizant strategy approaches zero almost
surely, for all initial states $x$ of player $i$.  Thus,
in particular, $\mu$ is approximately optimal for player $i$ in a
large finite game.  A weaker version of this property, called
the {\em approximate Markov equilibrium property}, was 
introduced by \cite{weintraub_2008}; a similar notion is also studied by
\cite{glynn_2004, huang_2005, tembine_2009} and \cite{bodoh_creed_2010}.

In order for this approximation property to hold, the key requirement
is that the model primitives $\pi(x,a,f)$ and $\mbf{P}(\cdot | x,
a,f)$ must be {\em jointly continuous} in $a$ and $f$, and the payoff
function must be uniformly bounded.  The intuition is that,
essentially, the desired approximation property amounts to a
continuity property in the value function of a player.  We refer the
reader to our companion paper \cite{adlakha_2010} for details of this
type of result in the case of discrete state spaces.  Independently of
our own work, \cite{bodoh_creed_2010} has also derived similar
conditions to ensure that mean field equilibrium approximates Markov
perfect equilibrium well, over compact continuous
state spaces.

For the remainder of the paper, we only study stochastic games
$\Gamma$ in the limiting regime where the number of players grows
large.  In particular, we focus on existence of, and convergence to,
mean field equilibrium.  In Section~\ref{sec:existence-cd}, we establish that a mean field 
equilibrium always exists for stochastic games with complementarities.

\subsection{Lattice-Theoretic Preliminaries}
\label{ssec:prelims}

This section contains an overview of some basic definitions and
notation used in the remainder of the paper. Our development requires some basic concepts from the theory of
lattices. Given a partially ordered set $X$ with order $\succeq$, an element $x$ is called an {\em upper
  bound} of $S$ if $x \succeq y$ for all $y \in S$; similarly, $x$ is
  called a lower bound of $S$ if $y \succeq x$ for all $y \in S$.  We
  say that $x$ is a {\em supremum} or {\em least upper bound} of $S$
  in $X$ if $x$ is an upper bound of $S$, and for any other upper
  bound $x'$ of $S$, we have $x' \succeq x$.  In this case we write $x = \sup S$.  We
  similarly define infimum (or greatest lower bound), and denote it by
  $\inf S$. The partially ordered set $(X, \succeq)$ is a {\em lattice} if for all
pairs $x, y \in X$, the elements $\sup\{ x, y\}$ and $\inf \{ x, y \}$
exist in $X$.
%We write $x \vee y = \sup\{ x, y\}$, and call this the
%{\em join} of $x$ and $y$; and we write $x \wedge y = \inf\{ x, y\}$,
%and call this the {\em meet} of $x$ and $y$.
The lattice $(X, \succeq)$ is a {\em complete lattice} if in addition, for all
nonempty subsets $S \subset X$, the elements $\sup S$ and $\inf S$
exist in $X$.
%A set $S$ is a {\em sublattice} of $(X, \succeq)$ if
%for any two $x, y \in S$, the elements $\sup_X \{x, y\}$ and $\inf_X
%\{x, y\}$ lie in $S$.  Note that $(S, \succeq)$ can be a lattice
%without being a sublattice; i.e., $\sup_S \{x, y\}$ and $\inf_S \{x,
%y\}$ may exist in $S$, but $\sup_X \{x, y\}$ and $\inf_X \{x, y\}$ may
%not lie in $S$.

If $X$ is a lattice, a function $f: \mc{X} \to \reals$ is {\em supermodular} if $f(\sup \{ x,x'\}) + f(\inf\{x,x'\}) \geq
f(x) + f(x')$ for every pair $x,x'$.  If $Y$ is also a lattice, a
function $f: X \times Y \to \reals$ has {\em increasing differences} in $x$
and $y$ if for all $x' \succeq x$, $y' \succeq y$, there holds
$f(x',y') - f(x',y) \geq f(x,y') - f(x,y)$.  Finally, a correspondence $T: 
X \to Y$ is {\em nondecreasing} if 
  whenever $x' \succeq x$, $y \in T(x)$, and $y' \in T(x')$, there holds
  $\sup \{ y,y'\} \in T(x')$, and $\inf \{ y, y'\} \in T(x)$.  
(For more detail on lattice programming, the reader is referred to \cite{topkis_1998}.)

Throughout this paper, we view $\mc{X}$ and $\mc{A}$ as
lattices in the usual ordering; since these spaces are both compact
subsets of $\reals$, the corresponding lattices are complete \citep{topkis_1998}.
We also view the set of strategies~$\mfr{M}_{O}$ as a lattice, under the coordinate
ordering $\unrhd$: i.e., $\mu' \unrhd \mu$ if and only if
$\mu'(x) \geq \mu(x)$ for all $x$.

In addition, recall that we let $\mfr{F}$ denote the 
set of all Borel probability measures on $\mc{X}$.
%\begin{align}
%\label{eqn:dist-set}
%\mfr{F} = \big\{f : \mc{X} \rightarrow [0, 1]\ |\ f(x) \geq 0, \sum_{x \in \mc{X}} f(x) = 1\big\}.
%\end{align}
We view $\mfr{F}$ as a lattice with the {\em (first order) stochastic
  dominance} ordering; formally, we write $f' \succeq_{\SD} f$ if and only if:
\[
\int_{x \in \mc{X}} g(x) f'(dx) \geq \int_{x \in \mc{X}} g(x)
f(dx) \]
for all nondecreasing, bounded, measurable functions $g$ on $\mc{X}$
(where the integral is the 
Riemann-Stieltjes integral).  It is straightforward to show that this condition is
equivalent to $F(x) \leq F'(x)$, where $F$ (resp., $F'$) is the
cumulative distribution function of $f$ (resp., $f'$).  It is
well known that $\mfr{F}$ is a lattice: the lattice supremum
$\sup_{\SD}\{ f, f'\}$ (resp., 
the lattice infimum $\inf_{\SD} \{ f, f'\}$) is found
by the pointwise infimum (resp., supremum) of the corresponding
distribution functions.  Because $\mc{X}$ is compact, it is
straightforward using an analogous argument to verify that $\mfr{F}$
is a complete lattice \citep{echenique_2003}.  

We conclude by defining some properties of parameterized distributions
we require in the sequel.  Let $f(\cdot|y)$ denote a family of
measures in $\mfr{F}$, parameterized by $y \in Y$, where $Y$ is a
lattice.  Then we say $f$ is {\em stochastically
  nondecreasing} in $y$ if whenever $y'$ is larger than $y$, $f(\cdot |
y') \succeq_{\SD} f(\cdot | y)$.  Similarly, let $f(\cdot | y,z)$
denote a family of measures in $\mfr{F}$ parameterized by $y \in Y$
and $z \in Z$, where both $Y$ and $Z$ are lattices.
Then we say that $f$ has {\em stochastically increasing differences}
in $y$ and $z$ if the expectation $\int_{x \in \mc{X}} g(x) f(dx | y,z)$
has increasing differences in $y$ and $z$, for every nondecreasing,
bounded, measurable function $g$ on $\mc{X}$.  

%%%%%

%% file: existence_cd.tex
\section{Existence of Mean Field Equilibria}
\label{sec:existence-cd}

In this section and the following section, we consider a baseline
model of stochastic games with complementarities, in which we prove
existence and convergence results.  In this section we establish our first main result: that there exists
a mean field equilibrium for the stochastic game
with complementarities. We also show an ordering result: there exists
a ``largest'' and a ``smallest'' equilibrium among the set of all mean field equilibria
with nondecreasing strategies.

We have the following definition.

\begin{definition}
\label{def:basic-cd}
A {\em stochastic game with complementarities} is a stochastic game
$\Gamma = (\mc{X}, \mc{A}, \P,
\pi, \beta)$ that satisfies the following properties.
\begin{enumerate}
\item {\em Nondecreasing and supermodular payoff.} The payoff $\pi(x, a, f)$ is nondecreasing in $x$, continuous in~$a$, and supermodular in $(x, a)$. Furthermore, for fixed $a$ and~$f$, $\sup_{x \in \mc{X}}\pi(x, a, f) < \infty$.

\item {\em Payoff complementarity.} The payoff function $\pi(x, a, f)$ has increasing differences in $(x, a)$ and~$f$.

\item {\em Monotone and supermodular transition kernel.} The transition
  kernel $\P(\cdot | x,a ,f)$ is stochastically supermodular in $(x, a)$ and
  is stochastically nondecreasing in each of $x$, $a$, and~$f$. Further, $\P(\cdot | x,a ,f)$ is continuous in $a$ (w.r.t the topology of weak convergence on $\mfr{F}$).

\item {\em Transition kernel complementarity.} The transition kernel
  $\P(\cdot | x,a ,f)$ has
  stochastically increasing differences in $(x, a)$ and~$f$.

\item {\em Monotone action set.}  The correspondence $\mc{A}(x)$ is
  nondecreasing in $x$.  \newrj{Further, $\sup_{a \in
      \mc{A}(x)} \pi(x,a,f)$ is nondecreasing in $x$ for all fixed $f$.}

\item {\em Countable noise.} For each $x$, $a$, and~$f$, the support $\{ x'\
: \ \P\left(x'  |  x, a, f\right) > 0 \}$ is countable. 
\end{enumerate}
\end{definition}

The first assumption is natural for a range of models---if larger
states are more valuable, then the payoff function will be 
nondecreasing in the state. The boundedness
assumption on the payoff will be trivially 
satisfied if, e.g., $\mc{X}$ is an interval and the payoff is
continuous in $x$. The second assumption ensures that there are
complementarities between the state and action of a single player and
the population state of other players. The next three assumptions
create complementarities between state and action, as well as ensure
that larger states and/or larger actions now are more likely to lead
to larger states in the future. The last assumption is made to
simplify later dynamic programming arguments; in particular, it allows
us to ignore measurability issues when considering optimal strategies
\citep{Bertsekas78}.  We note that if the payoff and transition kernel
are continuous, then countability becomes unnecessary for our
analysis, since we can restrict attention to optimal strategies that
are continuous in the state.

While it may be straightforward to verify whether a payoff function
exhibits the desired complementarity properties, the same verification
is somewhat more challenging for the transition kernel.  Thus before
continuing, we provide an example of a transition kernel that 
exhibits the complementarity conditions required in Definition
\ref{def:basic-cd}.

\begin{example}[Mixture dynamics]
Suppose that $\P(\cdot | x,a,f)$ is defined as follows:
\begin{equation}
\label{eq:mixture}
\P(\cdot | x,a,f) = q(x,a,f) F(\cdot) + (1 - q(x,a,f)) G(\cdot).
\end{equation}
Here $F$ and $G$ are both distributions on $\mc{X}$, such that $F$
first order stochastically dominates $G$, and $0 \leq q(x,a,f) \leq 1$.  If $q(x,a,f)$ is
nondecreasing in $x$, $a$, and $f$, supermodular in $(x,a)$, and has
increasing differences in $(x,a)$ and $f$, then it can be checked that
the expectation of \eqref{eq:mixture} against any nondecreasing
function satisfies all the conditions of Definition
\ref{def:basic-cd}.  As one example of a $q$ that satisfies these
properties, suppose:
\[ q(x,a,f) = \frac{x + a + \eta(f)}{2\sup \mc{X} + \sup \mathsf{A}}, \]
where $\eta(f) = \int_{\mc{X}} x' f(dx')$ is the mean of $f$.  
Such dynamics are commonly used in the context of games with strategic
complementarities~\citep{Curtat96}.
\hfill \halmos \end{example}

Informally, how might we expect players to behave in such a game?
Observe that if other players have a larger population state, this increases the
return to a larger state for a given player.  In order to achieve a
larger state, a player must take a larger action; but this also
increases the likelihood of larger states in the future.  All these
effects conspire to create a situation where, when players are
confronted with larger population states, they are likely to take
higher actions.  This monotonicity drives our analysis.  

For the remainder of the section we fix a stochastic game with
complementarities $\Gamma = (\mc{X}, \mc{A}, \P,
\pi, \beta)$.  Let $\Phi: 
\mfr{F} \to \mfr{F}$ denote the composition of $\mc{P}$ and $\mc{D}$
for the game $\Gamma$:
$\Phi(f) = \mc{D}(\mc{P}(f), f)$.  A fixed point of $\Phi$ identifies a
mean field equilibrium of $\Gamma$.  Intuitively, under the assumptions we have
made we might expect $\Phi$ to be a {\em monotone} map; i.e., larger
initial conjectures about the population state should lead players to
take higher actions, which should in turn lead to a larger invariant
distribution.  Tarski's fixed point theorem ensures monotone functions
on a lattice have a fixed  point.\footnote{Note that although Tarski's
  theorem applies to functions, in our case $\Phi$ is a
  correspondence. \cite{Zhou94} provides a generalization of Tarski's
  theorem to correspondences.}

\begin{theorem}[\citealp{tarski_1955}]
\label{th:tarski}
Suppose that $\mc{L}$ is a nonempty complete lattice, and $T : \mc{L}
\to \mc{L}$ is a nondecreasing function.  Then the set of fixed points
of $T$ is a nonempty complete lattice.
\end{theorem}

  We proceed to show that $\Phi$ is
monotone by showing that each of two 
correspondences $\mc{P}$ and $\mc{D}$ are 
monotone (with respect to the coordinate ordering on strategies in
$\mfr{M}_O$, and the first order stochastic dominance ordering on $\mfr{F}$).

Our main result in this section is the following theorem.

\begin{theorem}
\label{th:existence-cd}
There exists a mean field equilibrium for the stochastic game with
complementarities~$\Gamma$.
\end{theorem}

In the next section, we sketch a proof of this theorem; and in Section
\ref{ssec:ordering-cd}, we show that if we restrict attention to
equilibria where the strategy is nondecreasing, then there exists a
``largest'' equilibrium and a ``smallest'' equilibrium.

\subsection{Theorem \ref{th:existence-cd}: Proof Sketch}
\label{sec:existthmproof}

We sketch the proof of Theorem \ref{th:existence-cd}; each step is
filled in by the lemmas in the appendix.

\begin{enumerate}
\item[Step 1.] We show $\mc{P}(f)$ is nonempty, and that
  optimal strategies can be identified via Bellman's equation (Lemma
  \ref{lem:bellman-cd}). 
\item[Step 2.]  We show that the value function $V^*(x|f)$ is
  nondecreasing in $x$ and has
  increasing differences in $x$ and $f$.  We use this fact to show that:
\[ \pi(x,a, f) + \beta \int_{\mc{X}} V^*(x'|f) \P(dx' |x, a, f) \]
is supermodular in $(x,a)$ and has increasing differences in $(x,a)$ and $f$ (Lemmas
\ref{lem:Tincrdiff-cd}, \ref{lem:DPincrdiff-cd}, and \ref{lem:V*incrdiff-cd}).
\item[Step 3.]  We use the complementarity properties of the previous
  step to show that the strategies $\ol{p}(f)$ and $\ul{p}(f)$ are 
nondecreasing in the state $x$, where:
\begin{align} 
\ol{p}(f) &= \sup \mc{P}(f); \text{ and } \label{eq:ulpolp-cd}\\
\ul{p}(f) &= \inf \mc{P}(f). \notag
\end{align}
We also show that $\ol{p}$ and $\ul{p}$ are nondecreasing in $f$.
(These facts are shown in Lemma \ref{lem:Gammaincr-cd}).\footnote{See
  also \cite{Hopenhayn92, topkis_1998} and \cite{smith_2002} for other conditions
  that yield monotonicity of optimal solutions to dynamic programs.}
\item[Step 4.]  We show that when restricted to strategies $\mu$
  that are nondecreasing in state, $\ol{d}(\mu, f)$ and
  $\ul{d}(\mu, f)$ are nondecreasing in $\mu$ and $f$, where:
\begin{align}
\ol{d}(\mu, f) &= \sup \mc{D}(\mu, f); \text{ and } \label{eq:uldold-cd}\\
\ul{d}(\mu, f) &= \inf \mc{D}(\mu, f).\notag
\end{align} 
(This is shown in Lemmas \ref{lem:Qgamma-cd} and \ref{lem:Dgamma-cd}).
\item[Step 5.]  We conclude that the functions $\ol{\Phi}(f)$ and $\ul{\Phi}(f)$ are
nondecreasing in $f$, where:
\begin{equation}
\ol{\Phi}(f) = \ol{d}(\ol{p}(f), f); \ \ \ul{\Phi}(f) =
\ul{d}(\ul{p}(f), f).  \label{eq:ulphiolphi-cd}
\end{equation}
Thus both $\ol{\Phi}(f)$ and $\ul{\Phi}(f)$ possess fixed
  points by Tarski's theorem (Lemma \ref{lem:fixedpts-cd}).  These fixed
  points identify mean field equilibria.
\end{enumerate}

\subsection{Largest and Smallest Equilibria}
\label{ssec:ordering-cd}

Typically in games with supermodular structure, it is possible to show
various ordering relationships among the equilibria.  In particular,
there is typically a ``largest'' and ``smallest'' equilibrium \citep{Milgrom90}.  In our
setting, we might conjecture that the largest fixed point of
$\ol{\Phi}$ (resp., the smallest fixed point of $\ul{\Phi}$) is the
largest (resp., the smallest) mean field equilibrium of the stochastic
game $\Gamma$.  However, this need not be the
case: as seen above, monotonicity properties of the map $\mc{D}$ are
only inferred on the subset of strategies that are {\em nondecreasing}
in the state.  In general, such monotonicity properties might not
hold over the entire strategy set---i.e., $\ul{d}(\mu, g)$ and
$\ol{d}(\mu, g)$ may not be nondecreasing over the entire set
$\mfr{M}_{O}$.  These monotonicity properties are necessary for
establishing the ordering of equilibria in classical supermodular game
theory.

From the discussion in the preceding paragraph, however, observe that
if we restrict attention to nondecreasing strategies, then indeed an ordering result can be proven.  In particular, the following
corollary shows that any mean field equilibrium where the strategy is nondecreasing is
bounded above by the largest fixed point of $\ol{\Phi}$, and bounded
below by the smallest fixed point of $\ul{\Phi}$.  

\begin{corollary}
\label{cor:ordmfe-cd}
Let $\ol{f}$ be the largest fixed point of $\ol{\Phi}$, and let
$\ul{f}$ be the smallest fixed point of $\ul{\Phi}$, i.e.:
\begin{equation}
\label{eq:ulfolf-cd}
\ol{f} = \sup\{ f : \ol{\Phi}(f) = f \}; \ \ \ul{f} = \inf\{ f :
\ul{\Phi}(f) = f \}.
\end{equation}
Let $(\mu,
f)$ be any mean field equilibrium of the stochastic game with
complementarities $\Gamma$, where $\mu$ is nondecreasing.  Then $\ul{f}
\preceq_{\SD} f \preceq_{\SD} \ol{f}$, and thus $\ul{p}(\ul{f}) \unlhd
\mu \unlhd \ol{p}(\ol{f})$.
\end{corollary}

%%% \begin{proof}
%%% Existence of an MFE follows from Lemma \ref{lem:fixedpts}.  Suppose
%%% that $(\gamma, f)$ is any MFE where $\gamma$ is nondecreasing.  Then
%%% by definition of $\ul{p}$ and $\ol{p}$, we have $\ol{p}(f) \unrhd
%%% \gamma \unrhd \ul{p}(f)$.  Applying Lemma \ref{lem:Dgamma_mon}, we
%%% have $\ol{\Phi}(f) \succeq_{\SD} f \succeq_{\SD} \ul{\Phi}(f)$.  
%%% \end{proof}

%% file: convergence.tex
\section{Convergence to Equilibrium}
\label{sec:convergence}

In this section we show that a mean field equilibrium can be obtained using a natural form
of learning dynamics among the players.  We start by
considering a simple form of best response dynamics to compute equilibria, where
we iteratively apply the maps $\ol{\Phi}$ and $\ul{\Phi}$ defined in
\eqref{eq:ulphiolphi-cd}.  We argue that this process is
unsatisfactory, both from a computational and modeling standpoint,
and instead propose an alternate process we refer to as {\em myopic
  learning dynamics}; these dynamics are both computationally
simpler and correspond to a natural learning behavior among the
agents.  We show that this process converges to mean field
equilibria.

We fix a stochastic game with complementarities $\Gamma = (\mc{X},
\mc{A}, \mbf{P}, \pi, \beta)$.  Throughout this section we study
$\Gamma$ in the limit of a continuum of agents, consistent with our
definition of mean field equilibrium.

\subsection{Best Response Dynamics}
\label{ssec:conv-brd}

We start by considering the
following algorithm.\\

\noindent {\sc Algorithm L-BRD}:
\begin{enumerate}
\item Initialize the state of every agent to $\ul{x} = \inf \mc{X}$, and let $f_0$
  denote the resulting population state---i.e., $f_0$ places all its
  mass on $\ul{x}$.
\item At time $t$, let $\mu_{t+1} = \ul{p}(f_t)$, and let $f_{t+1} =
  \ul{d}(\mu_t, f_t)$, cf. \eqref{eq:ulpolp-cd} and \eqref{eq:uldold-cd}.  
\item Repeat (2).
\end{enumerate}\mbox{}\\

Here L-BRD denotes {\em lower best response dynamics}.  Given a
current population state, we compute the lowest best response of a player,
and then compute the smallest invariant distribution corresponding to
the resulting strategy.  This is the
simplest dynamic we might consider; since $\ul{\Phi}(f) =
\ul{d}(\ul{p}(f), f)$, we have $f_{t+1} = \ul{\Phi}(f_t)$.  In spirit, this
algorithm is similar to other best response dynamics that are common
in the literature on supermodular games \citep{Milgrom90, Vives90}.  

We now show that this algorithm converges; and further, under an
appropriate continuity condition, the limit point is the smallest mean
field equilibrium.  We have the following assumption.

\begin{assumption}
\label{as:continuity-cd}
The payoff function $\pi(x,a, f)$  and the transition probability measure $\mbf{P}(\cdot | x, a, f)$ are
both jointly continuous in their domains (where we endow $\mfr{F}$
with the topology of weak convergence).
\end{assumption}

The next proposition shows L-BRD converges; the proof follows by
exploiting monotonicity of~$\ul{\Phi}$. 

\begin{proposition}
\label{prop:conv-l-brd}
Let $\Gamma$ be a stochastic game with complementarities.  Define $f_t$ and $\mu_t$ iteratively according to Algorithm L-BRD.  Then $f_0 \preceq_{\SD} f_1
\preceq_{\SD} f_2 \cdots$, and $\mu_0 \unlhd \mu_1 \unlhd
\mu_2 \cdots$.  Further, there exists a distribution $f^*$ and
a strategy $\mu^*$, nondecreasing in $x$, such that $f_t$ converges
weakly to $f^*$ as $t 
\to \infty$, and $\mu_t$ converges pointwise to $\mu^*$ as $t \to
\infty$.

If, in addition, Assumption \ref{as:continuity-cd} holds, then 
$(\mu^*, f^*)$ is a mean field equilibrium, and $f^*$ 
is $\ul{f}$, the smallest fixed point of $\ul{\Phi}$
(cf. \eqref{eq:ulfolf-cd}).
\end{proposition} 

\newsa{Thus under mild continuity conditions on the model primitives,
  best response dynamics converge to a mean field
equilibrium.  Further, the limit point is the smallest mean field
equilibrium among all those where the equilibrium strategy is
nondecreasing.}

We conclude by noting that we can analogously define an {\em upper} best response
dynamic as follows.\\

\noindent {\sc Algorithm U-BRD}:
\begin{enumerate}
\item Initialize the state of every agent to $\ol{x} = \sup \mc{X}$; let $f_0$
  denote the resulting population state---i.e., $f_0$ places all its
  mass on $\ol{x}$.
\item At time $t$, let $\mu_{t+1} = \ol{p}(f_t)$, and let $f_{t+1} =
  \ol{d}(\mu_t, f_t)$, cf. \eqref{eq:ulpolp-cd} and \eqref{eq:uldold-cd}.  
\item Repeat (2).
\end{enumerate}\mbox{}\\

The same conclusion as Proposition~\ref{prop:conv-l-brd} holds for
U-BRD as well, except that under Assumption~\ref{as:continuity-cd}, the
limit point is the {\em largest} fixed point of $\ol{\Phi}$, i.e., $f^* =
\ol{f}$ (cf. \eqref{eq:ulfolf-cd}).

\newrj{We note that one alternative to L-BRD and U-BRD is
presented by \cite{sleet_2001}.  He suggests an algorithm based on
iterative value and policy iteration to compute a mean field
equilibrium of a dynamic price-setting game with stochastic,
exogenous firm-specific demand shocks per period.  The setting
considered there is specialized, but the convergence proof also
exploits monotonicity properties induced by complementarity
conditions in that specific model.}

\subsection{Myopic Learning Dynamics}

The preceding section establishes the desirable result that best
response dynamics converge.  However, in a dynamic context, iterative
application of $\ol{\Phi}$ and $\ul{\Phi}$ is not completely
satisfactory, whether viewed from a computational or modeling standpoint.
First, given $f$, computing $\ol{\Phi}(f)$ or $\ul{\Phi}(f)$ requires
computing the invariant distribution of the Markov chain induced by
$\ol{p}(f)$ or $\ul{p}(f)$, introducing additional complexity.
Second, the process of iteratively applying $\ol{\Phi}$ or $\ul{\Phi}$
does not naturally correspond to any reasonable dynamic process that
agents are likely to follow in practice: it is difficult to imagine an
agent first computing the invariant distribution of the current strategy
in use by her competitors, and then solving a dynamic program given that
invariant distribution.

By contrast, in this section we present a pair of {\em myopic learning
  dynamics} that address these considerations.  The algorithms
presented in this section are simple and easy to
implement. Furthermore, they demand only a weak form of rationality
from the players, thereby resolving the two main issues of {\em
  computability} and {\em plausibility} associated with the standard
solution concept of Markov perfect equilibrium (as discussed in the
Introduction).

In the myopic learning dynamic, at each time $t$, each agent computes
a best response to the current population state distribution $f_t$,
{\em assuming that the population state will remain at $f_t$ at all
  future times}.  (This step is similar to model predictive control or
receding horizon control; see, e.g., \cite{garcia_1989}.)  In other
words, agents play according to a strategy in $\mc{P}(f_t)$.  This
play yields a new population state $f_{t+1}$ at the next time step
according to the transition kernel.

The algorithms we consider are reasonable in settings where agents are
not likely to predict future learning by other agents.  Indeed, such
an assumption seems plausible precisely in the large systems that mean
field equilibrium
is meant to model.  In such systems, myopic behavior is simple
computationally; by contrast, solving a dynamic program with full
knowledge of future strategies other agents will employ places
unreasonable informational requirements on the agents.  

We first consider an algorithm where agents play actions induced by
$\ul{p}$.\\

\noindent {\sc Algorithm L-MLD}:
\begin{enumerate}
\item Every agent initializes their state to $\ul{x} = \inf \mc{X}$ at time $t = 0$.
\item Agents observe the population state $f_t$.
\item An agent with state $x$ chooses the action $a_t$ so that $a_{t} = \mu_{t}(x)$, where $\mu_{t}(x) =
  \ul{p}(f_t)(x)$.  The agent's next state is distributed according to
  $\P(\cdot | x, a_{t}, f_{t})$.
\item Repeat (2)-(3).
\end{enumerate}\mbox{}

Here L-MLD denotes {\em lower myopic learning dynamics}.  Observe that agents compute a new strategy based on the observed {\em
  current} population state---not based on the invariant distribution
associated to the last strategy chosen.  This means that two
simultaneous dynamic processes are taking place: strategy revision on
the part of the players, but also state update via the system dynamics
\eqref{eq:oe_dynamics}.  Due to this intertwined dynamic, novel
arguments are required to prove convergence of best response dynamics (relative to
usual proofs of convergence for such dynamics in supermodular games,
e.g., \citealp{Milgrom90, Vives90}).   We also note that although the same strategy is computed
by every agent, the particular action chosen will vary depending on their
current state.  

The preceding description yields a simple recursion
   for the population state at the next time step; for
  all Borel sets $S$:
\begin{equation}
\label{eq:brd-cd}
f_{t+1}(S) = \int_{\mc{X}} \P(S | x', \mu_{t}(x'), f_{t}) 
  f_t(dx') = Q_{\mu_{t}, f_{t}}(f_t)(S),
\end{equation}
where $Q_{\mu, f}(f)$ is defined as follows:
\begin{equation}
\label{eq:Qgamma-cd}
 Q_{\mu, f}(g)(S) = \int_{\mc{X}}\P(S|x, \mu(x), f)\; g(dx).
\end{equation}

Our goal is to understand the behavior of the sequence of population
states $f_0, f_1, f_2, \ldots$, as well as the sequence of policies
$\mu_0, \mu_1, \mu_2, \ldots$.   We have the following proposition,
which mirrors Proposition \ref{prop:conv-l-brd}.

\begin{proposition}
\label{prop:conv-cd}
Let $\Gamma$ be a stochastic game with complementarities.  Define $f_t$ and $\mu_t$ iteratively according to Algorithm L-MLD.  Then $f_0 \preceq_{\SD} f_1
\preceq_{\SD} f_2 \cdots$, and $\mu_0 \unlhd \mu_1 \unlhd
\mu_2 \cdots$.  Further, there exists a distribution $f^*$ and
a strategy $\mu^*$, nondecreasing in $x$, such that $f_t$ converges
weakly to $f^*$ as $t 
\to \infty$, and $\mu_t$ converges pointwise to $\mu^*$ as $t \to
\infty$.

If in addition Assumption \ref{as:continuity-cd} holds, then $(\mu^*, f^*)$ is a mean field equilibrium, and $f^*$
is $\ul{f}$, the smallest fixed point of $\ul{\Phi}$ (cf. \eqref{eq:ulfolf-cd}).
\end{proposition}

Thus we find the same result as for L-BRD: under mild continuity
conditions on the model primitives, the dynamics converge to the smallest mean field
equilibrium among all those where the equilibrium strategy is
nondecreasing.

The proof of Proposition \ref{prop:conv-cd} proceeds as follows.  We
exploit two key monotonicity properties established in the course of
proving existence of an equilibrium (Theorem \ref{th:existence-cd}):
first, that $\ul{p}(f)$ is monotone in $f$ (Lemma
\ref{lem:Gammaincr-cd} in the appendix); and second, that $Q_{\mu,
  f}(g)(S)$ is monotone in $\mu$, $f$, and $g$ (Lemma
\ref{lem:Qgamma-cd} in the appendix).  These two properties together
allow us to establish that $\mu_t$ and $f_t$ form monotone
sequences---even though players are reacting only to the current
population state, the population state over time moves monotonically
towards an equilibrium.

Note that L-MLD initializes players to the lowest state, $\inf \mc{X}$.  This
behavior of L-MLD is particularly meaningful for several of 
the applications described in the Introduction; for example, in an
interdependent security setting, we might envision a scenario where a
new, more efficient technology for security is introduced.  In this
case the ``low'' initial population state might correspond to the
status quo, and then the myopic learning dynamics track the adaptation
of the population to a new equilibrium configuration.

A similar convergence result also holds if instead every agent starts at
the {\em largest} state $\ol{x} = \sup \mc{X}$, and follows the strategy
$\ol{p}(f_t)$ at each time step.  We call this Algorithm U-MLD.\\

\noindent {\sc Algorithm U-MLD}:
\begin{enumerate}
\item Every agent initializes their state to $\ol{x} = \sup \mc{X}$ at time $t = 0$.
\item Agents observe the population state $f_t$.
\item An agent with state $x$ chooses the action $a_t$ so that $a_{t} = \mu_{t}(x)$, where $\mu_t(x) =
  \ol{p}(f_t)(x)$.  The agent's next state is distributed according to
  $\P(\cdot | x, \mu_{t}, f_{t})$.
\item Repeat (2)-(3).
\end{enumerate}\mbox{}

Note that \eqref{eq:brd-cd} continues to hold, with $\mu_t$ chosen
according to the preceding algorithm. The same conclusion as
Proposition~\ref{prop:conv-cd} holds for U-MLD as well, except that
under Assumption~\ref{as:continuity-cd}, the limit point is the {\em
  largest} fixed point of $\ol{\Phi}$, i.e., $f^* = \ol{f}$
(cf. \eqref{eq:ulfolf-cd}).

We conclude this section by discussing the behavior of myopic learning
dynamics in {\em finite} systems.  In particular, suppose that in a
game consisting of $m$ players, each player follows the dynamic
prescribed by L-MLD: each player starts in the lowest state, and then
at each time step, observes the current population state and plays one
step according to the optimal oblivious strategy given that population
state.  Because the system is finite, additional error is introduced
due to the randomness in state transitions of individual agents; in
particular, due to this randomness, it is not immediately guaranteed
that myopic learning dynamics will converge to a mean field
equilibrium in a finite game.  However, if the state space is
discrete, then using techniques similar to \cite{adlakha_2010} it can
be shown that $f_t^{(m)} \to f_t$ weakly, almost surely, where
$f_t^{(m)}$ is the population state after~$t$ time steps with $m$
players, and $f_t$ is the population state in the L-MLD dynamic after~$t$ time steps in the mean field limit.  Thus after sufficiently many
time steps and for sufficiently large
finite systems, the population state under L-MLD converges
approximately to a mean field equilibrium population state.  We
illustrate this point later in Section \ref{sec:numerical}.

%% file: comparative_statics.tex
\section{Comparative Statics}
\label{sec:comp-stat}

\newrj{In this section we discuss {\em sensitivity} analysis of equilibria,
also known as {\em comparative statics} results. Our goal is to
understand how the equilibrium distribution and optimal strategy are
altered in response to changes in parameters.  These results allow us to evaluate
changes in equilibrium with respect to changes in a parameter.}
%For
%example, these results allow to us answer the question of the form:
%{\em How does the equilibrium distribution and an optimal policy
%  changes when certain incentive mechanism change the structure of the
%  payoff function?}

In this section we consider a family of stochastic games with
complementarities, parameterized by $\theta \in \Theta$, where
$\Theta$ is a complete lattice. \newrj{In the context of security games, this
parameter could, for example, represent the effectiveness of a
particular security technology.  Alternatively, in the context of
recommendation systems, $\Theta$ might represent the effectiveness of
the collaborative filtering engine in improving recommendations to one
agent based on the profiles of other agents.}

Formally, suppose we are given a family of stochastic games
$\Gamma(\theta)$ for $\theta \in \Theta$ with common strategy spaces,
action spaces, and discount factors, where for each $\theta$,
$\Gamma(\theta)$ is a stochastic game with complementarities, i.e., $\Gamma(\theta)$
satisfies Definition~\ref{def:basic-cd} for each $\theta \in \Theta$. We refer
to $\Gamma$ as a {\em parametric family} of stochastic games with
complementarities.  Let $\pi(x,a,f;\theta)$ and $\mbf{P}(\cdot |
x,a,f;\theta)$ be the payoff and transition kernel, respectively, in
$\Gamma(\theta)$.  We make the following assumption.

\begin{assumption}
\label{as:comp-stat}
The payoff $\pi(x,a,f;\theta)$ has increasing differences in $(x,a,f)$
and $\theta$.  The transition kernel $\mbf{P}(\cdot |
x,a,f;\theta)$ has stochastically increasing differences in $(x,a,f)$
and $\theta$, and is stochastically nondecreasing in $\theta$ for fixed $x,a,f$.
\end{assumption}

Under the preceding assumption, we can give a directional
characterization of the movement of equilibrium in response to
parameter changes.  

\begin{theorem}
\label{th:comp-stat}
Let $\Gamma$ be
a parametric family of stochastic games with complementarities, and
suppose that Assumption \ref{as:comp-stat} holds.  Let 
$\ul{f}(\theta)$ and $\ol{f}(\theta)$ denote the ``smallest'' and
``largest'' equilibrium in the game $\Gamma(\theta)$,
cf. \eqref{eq:ulfolf-cd}.  Then $\ol{f}(\theta)$
and $\ul{f}(\theta)$ are both nondecreasing in $\theta$. 
\end{theorem}

Such comparative statics results are commonly applied in the
  context of games with complementarities; but it is worth noting that
  in a dynamic context this result provides additional insight, because
  it quantifies how the {\em distribution} of agents'
  states will respond as a parameter changes.  This kind of insight is
  particularly valuable for system designers, regulators, and policy
  makers, where changes in equilibrium behavior due to control
  decisions may be challenging to characterize.  As one simple
  consequence of the preceding theorem, suppose that in security
  games, an incentive is introduced for agents to invest in
  security as a linear rebate in the payoff, proportional to an
  agent's security level $x$.  It is straightforward to check that
  this results in more players opting for higher investment, and thus
  the equilibrium population state tends to shift towards higher security
  levels.

%% file: actions.tex
\section{Coupling Through Actions}
\label{sec:actions}

In the stochastic game model considered thus far, players' payoffs and
dynamics are ``coupled'' through their {\em states}; formally,
$\pi(x,a,f)$ and $\P(\cdot | x, a,f)$ depend on the population state
$f$, which is in turn a distribution over the state space $\mc{X}$.
In many models, however, the coupling between agents is through their
actions, rather than states; that is, $f$ is a distribution over the
{\em action} set $\mathsf{A}$, rather than over the state space.  Our
analysis extends rather easily to models of this form; in this section
we briefly discuss existence of, and convergence to, mean field
equilibrium in such models.

Formally, an {\em action-coupled stochastic game} $\Gamma = (\mc{X}, \mc{A}, \P,
\pi, \beta)$ has the following
distinctions from the (state-coupled) stochastic game defined in
Section \ref{sec:model}.

\textit{Population action distribution.}  We define the population
action distribution as
follows.  Let $\alpha_{-i,t}(a)$
denote the fraction of players (excluding player $i$) that play action
$a$ at time $t$, i.e.:
\begin{align}
\label{eqn:actual-dist-action}
\alpha_{-i,t}(a) = \frac{1}{m-1}\sum_{j \neq i}\indic{\{a_{j,t} = a\}},
\end{align}
where $\indic{\{a_{j,t} = a\}}$ is the indicator function that the
action of player $j$ at time $t$ is $a$.  We refer to $\alpha_{-i,t}$ as the
{\em population action distribution} at time $t$ (from player $i$'s point of view).

We let $\mfr{F}_{\mathsf{A}}$ denote all Borel probability measures
over $\mathsf{A}$.  Note that the population action distribution lies
in $\mfr{F}_{\mathsf{A}}$.  

\textit{Transition probabilities and payoff.}  We denote
the payoff by $\pi(x,a,\alpha)$, and the transition kernel by $\P(\cdot |
x, a,\alpha)$, where $\alpha$ is a population action distribution,
i.e., an element of $\mfr{F}_{\mathsf{A}}$.
\vspace{0.1in}

Recall that in defining mean field equilibrium in Section
\ref{subsec:OE}, we consider two maps $\mc{P}(f)$ and $\mc{D}(\mu,
f)$; the former gives the set of optimal oblivious strategies given a
population state $f$, and the latter gives the set of invariant
distributions under a kernel with strategy $\mu$ and population state
$f$.  Those maps are analogously defined for action-coupled stochastic
games, but with $f$ as the population action distribution rather than
the population state; we omit the formal details.  With a slight abuse
of notation, we let $\mc{P}(\alpha)$ be the set of optimal oblivious
strategies for a player, given population action distribution
$\alpha$; and we let $\mc{D}(\mu, \alpha)$ be the set of invariant
distributions of the dynamics induced by oblivious strategy $\mu$ and
population action distribution $\alpha$.

In order to define mean field equilibrium, we require one
additional function.  Given a population state $f$ and an oblivious strategy $\mu$, let
$\hat{\mc{D}}(\mu, f)$ give the resulting population action distribution;
i.e., for Borel sets $S$:
\[ \hat{\mc{D}}(\mu, f)(S) = \int_{\mu^{-1}(S)} f(dx). \]
Note that $\mu^{-1}(S)$ is the set of states $x$ such that $\mu(x)
\in S$.  In order for this definition to be well posed, we require
the strategy $\mu$ to be Borel measurable; to avoid this issue we simply
assume that all model primitives are continuous, i.e., that Assumption
\ref{as:continuity-cd} holds.  Under this assumption it can be shown that
we can restrict attention to Borel measurable strategies $\mu$.

If every agent conjectures that $\alpha$ is the long run population
action distribution, then
every agent would prefer to play an optimal oblivious strategy $\mu$.
On the other hand, if every agent plays $\mu$, and the long run
population action distribution is indeed $\alpha$, 
then $\alpha$ must also be the population action distribution that results from an
invariant distribution in $\mc{D}(\mu, \alpha)$.  This yields the
following definition of a mean field equilibrium for action-coupled stochastic games.

\begin{definition}
\label{def:oe-actions}
A strategy~$\mu$, population state~$f$, and population action
distribution $\alpha$ constitute a
mean field equilibrium of an action-coupled stochastic game $\Gamma$
if $\mu \in \mc{P}(\alpha)$,  $f \in \mc{D}(\mu, \alpha)$, and $\alpha
= \hat{\mc{D}}(\mu, f)$.  
\end{definition} 

An {\em action-coupled stochastic game with complementarities} is then defined
exactly as in Definition~\ref{def:basic-cd}, but with the population
state $f$ replaced by the population action distribution $\alpha$.
Extending the argument in the proof of Theorem
\ref{th:existence-cd}, we can prove the following theorem.

\begin{theorem}
\label{th:existence-action}
Suppose Assumption \ref{as:continuity-cd} holds.  Then there exists a
mean field equilibrium for any action-coupled stochastic game with 
complementarities $\Gamma$.
\end{theorem}

As is clear from the proof, the same monotonicity properties employed
to prove existence of a mean field equilibrium can also be used to
extend Corollary \ref{cor:ordmfe-cd} (establishing the existence of a
``largest'' and ``smallest'' mean field equilibrium) as well as
Proposition \ref{prop:conv-l-brd} and \ref{prop:conv-cd}
(establishing convergence of best response dynamics and myopic
learning dynamics, respectively).  We omit the details of these derivations as
they mirror earlier development in the paper nearly identically.

%% file: existence.tex
\section{Separable Stochastic Games}
\label{sec:existence}

%In this section, we study a particular class of stochastic games
%with complementarities where the dynamics are decoupled. In other words,
%the state evolution of a player~$i$ depends only on the past state and 
%action taken by the player at time~$t$. It is independent of the population
%state of other players. 

As the preceding sections illustrate, stochastic games with
complementarities possess a number of properties that make them
amenable to equilibrium analysis.  One potential concern, however, is that the set
of models admitted by Definition \ref{def:basic-cd} may be somewhat
limiting.  Consider the following example.

\begin{example}[Linear dynamics]
\label{ex:separable}
Consider a
simple model where the distribution of the next state of an agent is
``linear'' in $x$ and $a$.  Let $W$ be a zero mean random variable that takes countably
many values, and fix positive constants $A$ and $B$.  We consider a
state space $\mc{X} = [-\ul{M},\ol{M}]$, for some large positive
constants $\ul{M}, \ol{M}$; and let $\mc{A}(x) = [\ul{a}, \ol{a}]$ for
all $x$, where $\ul{a} \leq \ol{a}$.  Define $\P$
as follows:
\begin{equation}
\label{eq:linear} \P(x' | x, a) = \left \{
\begin{array}{ll}
\prob(Ax + Ba + W \geq \ol{M}),&\ \  x' = \ol{M};\\
\prob(Ax + Ba + W = x'),& \ \ -\ul{M} < x' < \ol{M};\\
\prob(Ax + Ba + W \leq -\ul{M}),& \ \ x' = -\ul{M}.
\end{array}
\right.
\end{equation}In this model, the state dynamics are essentially linear, except at
the boundaries of the state space (where the state is truncated to lie
within $[-\ul{M},\ol{M}]$).  Such a model might naturally arise in a wide range
of examples, e.g., Examples \ref{ex:ids}, \ref{ex:cf}, or \ref{ex:coord} (see
Section \ref{sec:examples} for details).

Unfortunately, such a kernel {\em does not} exhibit stochastically
increasing differences in general.  To see this, we consider a simple
instance where $\prob(W = 1) = \prob(W = -1) = 1/2$, and $\ul{M} =
\ol{M} = M \gg 1$.
Consider any nondecreasing function $\phi(x)$, and fix $x$ and $a$
such that $|Ax + Ba| < M -1$.  Then:
\[ \E[\phi(x') | x, a] = \frac{1}{2} \phi(Ax + Ba + 1) + \frac{1}{2}
\phi(Ax + Ba - 1). \]
In general, the right hand side exhibits increasing
differences in $x$ and $a$ only if $\phi$ is locally {\em convex}.  This is easiest
to see for differentiable $\phi$: in that case the cross partial
derivative $\partial^2 \phi(Ax + Ba + 1)/\partial x \partial a$ has to
be nonnegative to ensure increasing differences, which only holds if $\phi''(Ax + Ba + 1) \geq 0$.  For
general nondecreasing $\phi$, therefore, the expectation $\E[\phi(x')
| x, a]$ need not exhibit increasing differences in $x$ and $a$.  
\hfill \halmos \end{example}

The preceding example highlights a deficiency in stochastic games with
complementarities: while a rich class analytically, they do present
some restrictions from a modeling standpoint.  In this sense,
complementarity can appear to be a {\em brittle} property.

However, this same brittleness can actually become an advantage:
although at first glance it may appear that complementarity fails,
often simple transformations can lead to games that admit analysis via
complementarity methods even if the original game did not.  (A common
example is the class of {\em log-supermodular} games used extensively
in oligopoly theory, where the logarithm of the profit function may be
supermodular; see, e.g., \cite{Milgrom90} and \cite{Vives90} for details.)

In this
section we demonstrate that a wide range of models, including those
with dynamics similar to Example \ref{ex:separable}, can be
transformed to standard stochastic games with complementarities.
Further, the class of models we develop has the benefit that the
assumptions are typically easier to check in practice.  This
significantly widens the applicability of our theory to models where
the desired monotonicity properties may not be immediately apparent.

The class of games we consider in this section feature a payoff that
is {\em separable} in the state and action.  We have the following
definition.  

\begin{definition}
A {\em
  separable stochastic game} is a stochastic game $\Gamma = (\mc{X}, \mc{A}, \P,
\pi, \beta)$ with the following additional properties.

\begin{enumerate}
\item {\em Actions.}  There exist $\ul{a}, \ol{a}$, such
that $\mc{A}(x) = [ \ul{a}, \ol{a} ]$ for all $x$.
\item {\em Payoff.}  The single period payoff to player $i$ at time $t$
can be written as $\pi(x_{i,t}, a_{i,t}, \fmi) = v(x_{i,t}, \fmi) -
c(a_{i,t})$, where we refer to $v(x, f)$ as the {\em utility} at state
$x$ and population state $f$, and $c(a)$ as the {\em cost} for action
$a$.
\item {\em Transition probabilities}.  The state of a player 
evolves according to a Markov process with the following transition probabilities.
If the state of player $i$ 
at time $t$ is $x_{i,t} = x$ and the player takes an action $a_{i,t} =
a$ at time $t$, then the next state is distributed according to the
Borel probability measure $\P(\cdot| h(x, a), f)$, where for Borel sets $S \subset
\mc{X}$,
\begin{align}
%\label{eqn:tm}
\P(S| h(x, a), f) = \prob\left(x_{i,t+1} \in S  |  x_{i,t} = x,
a_{i,t} = a, \v{f}_{-i,t} = f\right).
\end{align}
Note that $\P$ depends on $x$ and $a$ only through the function
$h(x,a)$; we refer to $h(x,a)$ as the {\em kernel parameter}.  We assume that $h$ takes values in a compact interval $H
= [\ul{h}, \ol{h}] \subset \reals$.
\end{enumerate}
\end{definition}

%\textit{Transition Probabilities.} The state of a player evolves
%according to the following Markov process. If the state of player $i$

In this section we provide insight into {\em separable stochastic
  games with complementarities}.  We have the following definition.
\begin{definition}
\label{def:basic}
A {\em separable stochastic game with complementarities} is a
separable stochastic game $\Gamma = (\mc{X}, \mc{A}, \P,
\pi, \beta)$ with the following properties.
\begin{enumerate}
\item {\em Nondecreasing payoff and convex cost}. The utility function $v(x,f)$ 
is nondecreasing in $x$, and the cost function $c(a)$ is nondecreasing and convex in
  $a$.  Further, for fixed $f$, $\sup_{x \in 
    \mc{X}} | v(x,f)| < \infty$.  
\item {\em Payoff complementarity}.  The utility function $v(x,f)$ has
  increasing differences in $x$ and $f$.
\item {\em Monotone transition kernel}.  The transition
  kernel $\P(\cdot | \hat{h}, f)$ is stochastically
  nondecreasing in $\hat{h}$ and $f$.
%: if $\hat{h}'
%  \geq \hat{h}$, then $\P(\cdot | \hat{h}') \succeq_{\SD} \P(\cdot | \hat{h})$.
  Further $\P(\cdot | \hat{h}, f)$ is continuous in $\hat{h}$
  (w.r.t. the topology of weak convergence on $\mfr{F}$).
\item {\em Transition kernel complementarity.} The transition kernel
  $\P(\cdot | \hat{h}, f)$  has
  stochastically increasing differences in $\hat{h}$ and~$f$.
\item {\em Kernel parameter monotonicity and complementarity}.  The
  kernel function $h(x,a)$ is 
  supermodular in $x$ and $a$, nondecreasing in the state $x$, and
  concave and nondecreasing in the action~$a$. 
\item {\em Countable noise.}  For each $\hat{h}$, the support $\{ x'\ :\ \P( x' | \hat{h}) > 0\}$ is countable. 
\end{enumerate}
\end{definition}

%The first assumption is analogous to that made before; note however that the payoff is not required to be nondecreasing in state, The second assumption on the convexity of the cost of action is important to allow us to re-parameterize the problem.  The next three assumptions are natural analog of the assumptions previously made. Note that the dynamics considered in the example above do satisfy all of the assumptions made above. 

We proceed by
reparametrizing the strategy in terms of the kernel parameter; under
this reparametrization, the resulting model is revealed to be a
special case of the general model studied earlier in this paper.

Formally, suppose we are given a separable stochastic game with complementarities
$\Gamma = (\mc{X}, \mc{A}, \P,
\pi, \beta)$.   Before we proceed, we require some additional
notation.  For each $x$, define:
\begin{equation}
\label{eq:H}
  H(x) = \{ \hat{h} : h(x,a) = \hat{h} \text{ for some } a \in
  \mc{A}\}.
\end{equation}
Thus $H(x)$ is the image of $\mc{A}$ under $h(x, \cdot)$.  
In addition, for each $\hat{h} \in H(x)$, define:
\begin{equation}
\label{eq:C}
C(x,\hat{h}) = \inf_{a \in A : h(x,a) = \hat{h}} c(a).
\end{equation}
Thus $C(x,\hat{h})$ is the minimum cost incurred to achieve kernel parameter
$\hat{h}$ when at state $x$.

The next lemma establishes some basic properties of $H$ and $C$.
It uses the assumption that the cost function is a convex function of 
action~$a$.

\begin{lemma}
\label{lem:HCprops}
Suppose $\Gamma$ is a separable stochastic game with
complementarities.  Suppose $H(x)$ and $C(x,\hat{h})$ are defined as
in \eqref{eq:H} and \eqref{eq:C}, respectively.  Then for each $x$,
$H(x)$ is a compact interval, 
and the sets $H(x)$ are nondecreasing in $x$.  

The function $C(x,\hat{h})$ is convex and
nondecreasing in $\hat{h}$ on $H(x)$ for each $x$, and nonincreasing
in $x$ for each $\hat{h}$ as long as $\hat{h} \in H(x)$. \newrj{Further, for
all $x$:
\begin{equation}
\label{eq:Cmin}
 \inf_{\hat{h} \in H(x)} C(x, \hat{h}) = c(\ul{a}).
\end{equation}
}

 If $x' > x$,
$\hat{h}', \hat{h} \in H(x') \cap H(x)$, and $\hat{h}' > \hat{h}$,
then:
\[ C(x', \hat{h}') - C(x, \hat{h}') \leq C(x', \hat{h}) - C(x,
  \hat{h}). \]
In other words, $C(x,\hat{h})$ has decreasing differences in $x$ and $\hat{h}$.
\end{lemma}

We now use Lemma \ref{lem:HCprops} to define a {\em new} stochastic
game, which is in fact a stochastic game with complementarities as
in Definition \ref{def:basic-cd}.

\begin{proposition}
\label{prop:transform}
  Suppose that $\Gamma = (\mc{X}, \mc{A}, \P, \pi, \beta)$ is
  a separable stochastic game with complementarities.  Define a new game
  $\hat{\Gamma} = (\hat{\mc{X}}, \hat{\mc{A}}, \hat{\P},
  \hat{\pi}, \beta)$, where:
\begin{enumerate}
\item $\hat{\mc{X}} = \mc{X}$; 
\item $\hat{\mc{A}}(x) = H(x)$ for all $x \in \mc{X}$;
\item $\hat{\P}(x' | x, \hat{h}, f) = \P(x' | \hat{h}, f)$; and
\item $\hat{\pi}(x, \hat{h}, f) = v(x, f) - C(x, \hat{h})$,
\end{enumerate}
with $H(x)$ and $C(x, \hat{h})$ are defined in \eqref{eq:H} and
\eqref{eq:C}, respectively.  Then $\hat{\Gamma}$ is a stochastic game
with complementarities, cf. Definition \ref{def:basic-cd}.
\end{proposition}

Based on the preceding proposition we have the following theorem.

\begin{theorem}
\label{th:separable}
Any separable stochastic game with complementarities $\Gamma$ has a mean field
equilibrium.
\end{theorem}

The preceding result can be extended, of course, to provide analogs of
Corollary \ref{cor:ordmfe-cd} (existence of a largest and smallest
equilibrium), as well as Propositions \ref{prop:conv-l-brd} and
\ref{prop:conv-cd} (convergence of best response dynamics and myopic
learning dynamics, respectively).  (The appropriate generalization of
Assumption \ref{as:continuity-cd} is that $\mbf{P}$ should be jointly
continuous in $\hat{h}$ and $f$, and $h$ should be jointly continuous
in $x$ and $a$.)  Note, however, that the dynamics defined here are in
the {\em modified} strategy space, where the ``action'' is the kernel
parameter chosen.  In particular, the dynamics in the original action
space may not be monotone at all; nevertheless, the eventual limit
point is a mean field equilibrium.

It is also straightforward to
generalize the comparative statics result in Theorem
\ref{th:comp-stat} to separable stochastic games using the same
transformation as the preceding result.  In
addition, the definition of a separable stochastic game with
complementarities can be naturally extended to separable
action-coupled stochastic games with complementarities (simply by
replacing the population state $f$ by the population action
distribution $\alpha$ in the payoff and transition kernel), and an
argument similar to Proposition \ref{prop:transform} shows that such a
game can be transformed to a standard action-coupled stochastic game
with complementarities.  

We conclude this section by noting that the preceding results continue
to hold in a setting where the payoff is not necessarily monotone, as
long as dynamics are decoupled.  Formally, suppose that $\Gamma = (\mc{X}, \mc{A}, \P, \pi, \beta)$ is a stochastic game that
satisfies all the conditions in Definition \ref{def:basic}, {\em
  except} that~$v$ is not necessarily nondecreasing in $x$.  Suppose
in addition that $\P(\cdot | \hat{h},f)$ does not depend on $f$;
thus we denote the kernel simply $\P(\cdot | \hat{h})$.  In this
model it can again be shown that a mean field
equilibrium exists, as we now describe.

The proof of Theorem \ref{th:existence-cd} (and subsequent results on
ordering of equilibria and convergence) use the fact that the payoff
is nondecreasing in $x$ to show that $\int_{\mc{X}} V^*(x' | f)
\P(dx' | x, a, f)$ is supermodular in $(x,a)$ and has increasing
differences in $(x,a)$ and $f$ (see Lemma \ref{lem:Tincrdiff-cd}).  In
order for the expectation to preserve these properties, the integrand
must be nondecreasing in state; this is why we require the payoff to
be nondecreasing.  However, if $\P$ only depends on the kernel
parameter, then we can show that $\int_{\mc{X}} V^*(x' | f) \P(dx'
| \hat{h})$ has increasing differences in $x$ and $\hat{h}$, even if
the payoff is not necessarily nondecreasing.  For details, we refer
the reader to Lemma \ref{lem:Tincrdiff} in the Appendix.  Substitution
of this lemma in the proof of Theorem \ref{th:existence-cd} yields the
desired result.

%% file: examples.tex
\section{Examples}
\label{sec:examples}

In this section we revisit the five examples mentioned in the
Introduction: interdependent security; collaborative filtering;
dynamic search with learning; coordination games; and oligopolies with
complementarities.  We show that each of these examples can be
formalized within the framework developed in this paper, so that the
existence and convergence results we have proven apply.

\subsection{Example \ref{ex:ids}: Interdependent Security}
\label{subsec:ids}
We consider a dynamic model of interdependent security in a
computer cluster, where the state~$x$ gives the  
security level of a player. Players can improve their security level
through investment; an investment $a$ incurs a cost $c(a)$ that is
convex and nondecreasing in $a$.  A higher action leads to improvement in
the security level, and with no or little investment the security
level deteriorates due to depreciation.  Thus a reasonable model for
the dynamic evolution of the security level might be the linear
dynamics in \eqref{eq:linear}, where $\ul{M} = 0$ and $\ol{M} \gg 0$,
and $A = 1$, $B > 0$, and $W$ has a negative expected value.
%Stochasticity is
%introduced due to uncertainty 
%about the effectiveness of any particular investment.
Let $p(x)$ be
the probability of a bad event 
occurring when an individual computer is at the security level~$x$, and
let $L$ be the cost of this bad event to the host.

We consider a simplified model where at each time step, an
individual computer ``talks'' to a randomly selected computer in the
network. (This talk can be in form of establishing a TCP connection,
exchanging data, emails, etc.) Thus, at each time, there is a
probability that an individual computer will suffer a bad event
because of the security level of the rest of the network. Let
$f_{-i,t}(y)$ be the fraction of all computers (except computer~$i$)
that have their security level at $y$ at time $t$. Then, at each time step,
computer~$i$ receives an expected value that is given as: 
\[
v(x_{i,t}, \v{x}_{-i,t})  = -p(x_{i,t})L - \left(1- p(x_{i,t})\right)\left(\sum_{y}f_{-i,t}(y)p(y)\right).
\]
The first part of the payoff reflects the security of host $i$.  The
scaling factor $1 - p(x_{i,t})$ in the second term is the 
probability that no bad event happens because of the individual
security level. The term $\sum_{y}f_{-i,t}(y)p(y)$ represents the
average security level of the rest of the network.  Because $p$ is
decreasing, it is straightforward to verify that the product of
these two terms exhibits strategic complementarities between the
security level of agent $i$, and the security level of every other
agent.  It follows that this is a {\em separable stochastic game with
  complementarities}.  

\subsection{Example \ref{ex:cf}: Collaborative Filtering}
\label{subsec:netflix}
As a canonical example, we consider the collaborative filtering system
used by a recommendation engine on a movie rental site such as
Netflix.  We let the state $x$ be the quality of a user's
profile, and assume $x$ takes values in a compact interval. The action
$a$ represents the effort put forth in updating her profile, e.g.,
through rating more movies; actions are costly, with $c(a)$ denoting
the cost incurred by action $a$.  We assume $c$ is convex.  If user
$i$ does not put forth any effort 
at time~$t$, then the profile becomes ``stale,'' i.e., the quality of
the profile drops over time.  Thus in this model the quality can be
modeled via dynamics as in \eqref{eq:linear} as well, where $A = 1$, $B
> 0$, and $W$ has negative expected value.

Based on the quality of a user's profile $x$ as well as the profile of
other users in the system (captured by the population state $f$), the
recommendation system suggests a movie to a user.  Let $v(x,f)$
denote the expected desirability of the movie recommended to a user,
given their profile quality $x$ and the population state $f$.  Observe
that $v$ will increase if $x$ increases, since a more accurate
profile results in more accurate recommendations.  However, for most
collaborative filtering systems, it is also the case that {\em if
  others have higher quality profiles, then the marginal return to a
  higher quality profile is higher;} for example, this would be the
case under a nearest neighbor algorithm as is commonly used by a
variety of online recommendation systems.  Thus such a model is a
separable stochastic game with strategic complementarities.
Collaborative filtering systems are one example of a setting with {\em
  positive network effects}; games with strategic complementarities
are commonly used to model settings with positive network effects.

\subsection{Example \ref{ex:dsl}: Dynamic Search with Learning}
\label{subsec:search}

\newrj{We consider a model where at each time step, a
trader exerts effort to search for trading partners.  As discussed in Example \ref{ex:dsl} in
the Introduction,
traders' experience grows with both their own effort {\em and} the effort of
others.  To formalize this notion, suppose traders choose effort each
time step from $[0,\ol{a}]$, where $\ol{a} > 0$.  Let the state $x$
denote the current {\em search productivity} of a given trader; we
assume $x \in [0,\ol{x}]$ where $\ol{x} > 0$.  Finally, given a population action distribution $\alpha$,
we let $\eta(\alpha)$ be defined 
as:
\[ \eta(\alpha) = \int_0^{\ol{a}} a \alpha(da). \]

We then assume that traders receive a payoff $\pi(x,a,\alpha)$ defined
as:
\[ \pi(x,a,\alpha) = xa\eta(\alpha) - c(a), \]
where $c(a)$ is a cost of effort.  In particular, observe that
the first term of the payoff increases as the search productivity
increases, the players own effort increases, or the
mean effort of others in the system increases.  

As a trader exerts effort, they gain experience and their search
becomes more productive.  Further, as discussed in Example \ref{ex:dsl} in
the Introduction,
traders' experience grows with both their own effort {\em and} the effort of
others; as in our other examples, with insufficient effort the
search productivity decreases as previously acquired experience
becomes outdated.  Thus we assume the transition kernel is defined as
in \eqref{eq:mixture}, where:
\[  q(x,a,\alpha) = \frac{x + a + \eta(\alpha)}{\ol{x} + 2 \ol{a}}. \]

This is a model where traders are coupled through their
actions, cf. Section \ref{sec:actions}.  It is straightforward to
verify that this model exhibits the complementarity properties
required for an action-coupled stochastic game with complementarities.}

% Let~$x \in [\ul{x},
% \ol{x}]$ be the coefficient of search productivity for a trader. Note
% that it is natural to assume that $\ul{x} 
% \geq 0$ since the search productivity is typically nonnegative. At
% each time step, a trader exerts an effort~$a \in [0, \ol{a}]$ to
% search for trading partners.  As discussed in Example \ref{ex:dsl} in
% the Introduction,
% traders' experience grows with both their own effort {\em and} the effort of
% others.   In particular, we assume that the search
% productivity of a trader grows depending on her own current search
% productivity, the effort exerted by the trader, and the {\em mean}
% effort of all other traders.  Given a population action distribution $\alpha$,
% we let $\eta(\alpha)$ be defined 
% as:
% \[ \eta(\alpha) = \int_0^{\ol{a}} a \alpha(da). \]
% We then assume that the transition kernel is defined as in
% \eqref{eq:mixture}, where:
% \[  q(x,a,\alpha) = \frac{x + a + \eta(\alpha)}{\ol{x} + 2 \ol{a}}. \]
% We assume that traders receive a payoff $\pi(x,a,\alpha) = xa\eta(\alpha) -
% c(a)$, where $c(a)$ is a cost of effort.  Note that the trader's
% benefit is directly proportional to their own experience, their own
% effort, and the mean effort of others.
% Thus this is a model where traders are coupled through their
% actions, cf. Section \ref{sec:actions}.  It is straightforward to
% verify that this model exhibits the complementarity properties
% required for an action-coupled stochastic game with complementarities.

\subsection{Example \ref{ex:coord}: Coordination Games}
\label{subsec:coordination}
In this model, a collection of agents are interested in
coordinating on a common state; the model we present is a related to
the one studied by \cite{huang_2006}.  Actions can alter the state, but
any nonzero actions are costly. We assume that $\mc{X} = [-M,M]$, and $\mc{A} = [0,L]$.
We assume dynamics are linear, cf. \eqref{eq:linear}, where $A,B > 0$,
and $W$ has negative expected value.
Each player tries to
minimize mean squared error to the other players' average
state, and incurs a quadratic cost for taking nonzero action.  If $f$ is the current
population state, we let:
\[ \eta(f) = \int_{\mc{X}} x f(dx). \]
We assume the payoff of a player is:
\[ \pi(x,a,f) = -(x - \eta(f))^2 - a^2. \]
It is straightforward to verify that $-(x-\eta)^2$ has increasing differences in $x$ and $\eta$, since the
cross partial derivative with respect to $x$ and $\eta$ is positive
\citep{topkis_1998}.  It follows that $v(x,f) = -(x - \eta(f))^2$ has
increasing differences in $x$ and $f$.  Further, $c(a) = a^2$ is
convex and nondecreasing in $a \in [0,L]$.

In principle we would like to claim this is a separable stochastic
game with complementarities, but the payoff is not
monotonic in $x$.  As discussed at the end of Section
\ref{sec:existence}, however, if the transition kernel does not depend on $f$ (as
is the case in \eqref{eq:linear}), then the payoff need not be
monotonic in $x$---and all our results continue to hold.  Thus
existence of equilibrium and convergence of MLD can be guaranteed for
this model.  Notably, our convergence result provides justification
for a distributed control interpretation of this coordination game,
where multiple individual agents can execute a myopic algorithm and
yet converge to a common state.

\newrj{We conclude by noting one peculiarity of our formulation: we
  have $\mc{A} = [0,L]$, so in particular, players cannot move {\em
    backwards} (i.e., take negative action).  This assumption is made
  to ensure that $c(a)$ is nondecreasing, as required for a separable
  stochastic game with complementarities.  However, in the original
  formulation of \cite{huang_2006}, $W$ has zero expected value, but
  players are allowed to take both positive {\em and} negative
  actions.  This expanded formulation can still be analyzed using the
  methods of this paper.  

Formally, suppose that $W$ has zero expected value, and $\mc{A} = [-L,L]$.  Then even
  though $c(a) = a^2$ is no longer nondecreasing on $\mc{A}$, we can
  still show in this specific model that
  $C(x,\hat{h})$ (cf. \eqref{eq:C}) exhibits decreasing
  differences in $x$ and $\hat{h}$.  To see this, note that with the 
  linear dynamics of \eqref{eq:linear}, $C(x,\hat{h}) = (\hat{h} -
  Ax)^2/B^2 $ for $\hat{h} 
  \in H(x)$.  Upon differentiating it follows that $\partial^2
  C/\partial x \partial \hat{h} \leq 0$, establishing  that $C(x,\hat{h})$ has decreasing
  differences in $x$ and $\hat{h}$.  By substituting this observation in the analysis of
  Section \ref{sec:existence} we recover all the results of that
  section.}

\subsection{Example \ref{ex:cg}: Oligopolies and Complementary Goods}
\label{subsec:quantity-comp}
Consider an oligopoly scenario where the goods produced by firms are
complements.  As firms gain experience in production, their cost of
production decreases.  We let $x \in [0, \ol{x}]$ be the experience level of a firm,
and let $a \in [0, \ol{a}]$ be the quantity produced by a firm.  Let $P(a, \alpha)$ be the
inverse demand curve seen by a firm, where $\alpha$ is the population
action distribution.  Thus this is a monopolistic competition model, where
firms sell differentiated products and the market clearing price seen by a firm
depends on the quantities produced by all firms.  The per period
payoff to a firm is:
\[ \pi(x,a,\alpha) = a P(a, \alpha) - c(x,a), \]
where $c(x,a)$ is the cost of producing quantity $a$ when a firm's
experience level is $x$.  Note that since~$\alpha$ is the population action
distribution, this is a game with coupling through actions.

We assume that firms' experience levels increase with higher
quantities produced; for example, we might consider dynamics of the
form \eqref{eq:mixture} with $q(x,a,\alpha)$ defined as:
\[ q(x,a,\alpha) = \frac{x + a}{\ol{x} + \ol{a}}. \]
Note in particular that in this model experience levels evolve independently
across firms.  

We note that the cost of production will typically decrease with the
experience level. Thus, $c(x,\alpha)$ is decreasing in $x$.  Further, at a
higher experience level, a firm's marginal cost 
of production typically decreases, so we expect 
$c(x, a)$ to have decreasing differences in~$x$ and~$a$.  Finally, since the payoff is
separable in~$x$ and~$\alpha$, it has increasing differences in those two
parameters.  

Since goods are complements, if $\alpha' \succeq_{\SD} \alpha$, then we
expect for a fixed production quantity $a$ the price is higher at~$\alpha'$, i.e., $P(a, \alpha') \geq P(a, \alpha)$ for every fixed~$a$. Furthermore,
it is natural that if~$\alpha' \succeq \alpha$, then for a slight increase in
production, a firm can charge a higher price for its goods. In other words,
$P(a, \alpha)$ should have increasing differences in~$a$ and~$\alpha$. Under these
natural assumptions, it is straightforward to verify that~$\pi(x,a,\alpha)$
has increasing differences between~$a$ and~$\alpha$. Thus this game is a
action-coupled stochastic game with complementarities.

%% file: numerics.tex
\section{Numerical Analysis}
\label{sec:numerical}

In this section, we study a numerical example that highlights the
utility of mean field equilibrium as a tool to analyze large scale stochastic games with
complementarities. Specifically, we consider an interdependent
security model, cf. Example \ref{ex:ids} and Section \ref{subsec:ids}.
We have three main goals.  First, we illustrate that mean field equilibrium provides
basic structural insight into equilibria in a simple and computable
fashion; in particular, we provide comparative statics analysis with
respect to cost and transition kernel parameters.  Second, we use the
model to evaluate the effect of heterogeneous player populations on
the equilibrium outcome.  Finally, we also evaluate how well myopic
learning dynamics perform in systems with finitely many players.  Our
analysis suggests that, particularly for stochastic games with
complementarities, mean field equilibrium is a powerful analytical tool that provides
rich structural insights relatively painlessly to the modeler.

\subsection{Model}

We assume that the
security level of a player is a positive integer~$x \in [0, 50]$ with
the interpretation that a higher value of the state implies a higher
security level. The probability that a player does not get
infected is assumed to be {\em proportional} to its state; after
normalization we have:
\[
1- p(x) = \frac{\kappa x}{1+ \kappa x},
\]
where $\kappa > 0$ is a scaling factor. At each time step, a player
takes an integer action~$a \in [0, 25]$ to improve its security
level.\footnote{\newrj{Note that in separable stochastic games with
  complementarities as defined in Section \ref{sec:existence}, we
  require actions to be chosen from a continuous interval.  However,
  for computational purposes, we consider a discrete approximation to
  the model proposed there, where actions are drawn from a discrete set.}}
  This
action results in a cost $c a$, where $c > 0$ is the marginal cost of
action. The payoff also depends on the average security level of other
players in the system. For a fixed player $i$, we let $\eta_{-i} = \sum_{y}f_{-i,t}(y)p(y)$
denote the average security level of the system (from the viewpoint
of player~$i$). Thus, the per period payoff to player $i$ is given by: 
\[
\pi\left(x_i, a_i, f_{-i}\right) = -p(x_i) -(1- p(x_i))\eta_{-i} - c a_i.
\]
Here~$f$ is the population state.

For the purposes of this example, we restrict attention to a separable
stochastic game, where the dynamics depend on the kernel parameter
given by $h(x, a) = x + a$. At each time step, based on the kernel
parameter, the next step is stochastically distributed as follows: 
\[ \P(x' | h) = \left \{
\begin{array}{ll}
\prob(h + W \geq 50),&\ \  x' = 50;\\
\prob(h + W = x'),& \ \ 0 < x' < 50;\\
\prob(h+W \leq 0),& \ \ x' = 0.
\end{array}
\right. \]
Here~$W$ is a random variable that takes values in the discrete set
$\{-1, 0, 1\}$ with the probability mass function given by: 
\[
\prob(W = w) = \left \{
\begin{array}{cccc}
q_{-1}, & \ \ w &= &-1;\\
q_0, & \ \ w &= &0;\\
q_1, & \ \ w &= &1;
\end{array}
\right.
\]
We initially choose $q_{-1} = q_1 = 0.4$, and $q_0 = 0.2$.

A player maximizes its expected discounted payoff with the discount
factor ~$\beta = 0.75$. We compute the mean field equilibrium using
the L-MLD algorithm, where we use value iteration to compute an
optimal oblivious strategy given the current population state.  For the purposes
of this simulation, we declare value iteration to have converged
if the total difference (across all states) between iterates is less
than~$10^{-4}$. Having computed the optimal strategy
for a given population state, the next population
state is computed using the recursion given in
\eqref{eq:brd-cd}.  For each simulation scenario, we
run~$1000$ iterations; for reference, we note that each run takes 
approximately~$8-10$ minutes on an Intel Core 2 Quad Q6600 (2.4GHz)
machine with 3GB RAM. At the end
of~$1000$ iterations, the total variation distance between the current
population state and the previous population state is always less than~$5 \times
10^{-4}$, so we refer to the population state at $t = 1000$ as the mean field equilibrium 
population state.

\begin{figure}
\centering
\includegraphics[width=5.5in]{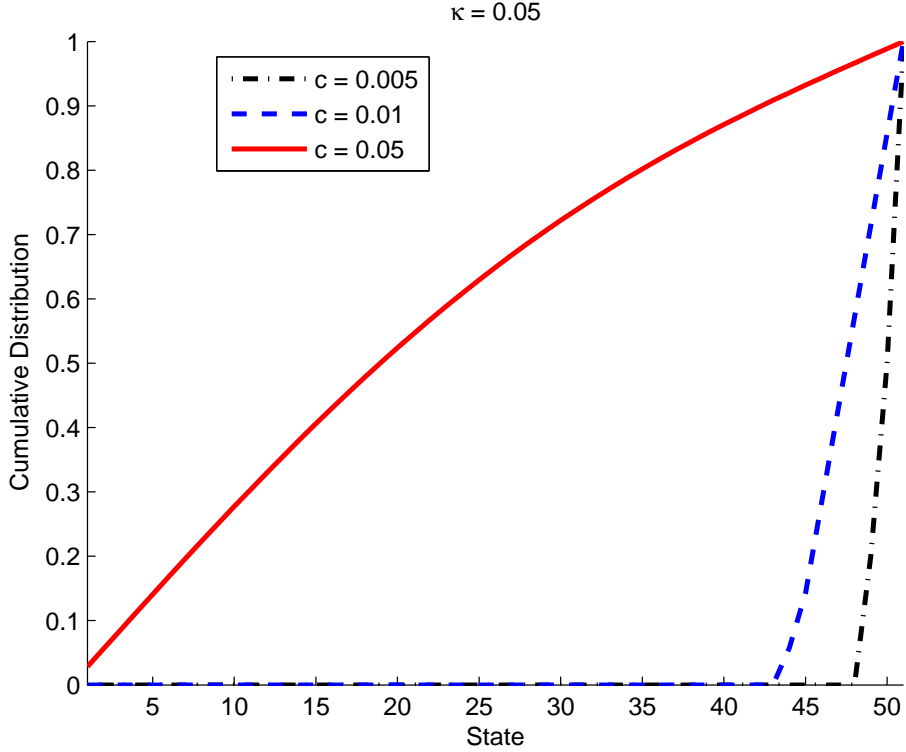}
\caption{Cumulative distribution function of population state at $t =
  1000$ under L-MLD for~$\kappa = 0.05$ and $c = 0.005, 0.01, 0.05$.}
\label{Fig:compStat1}
\end{figure}

\subsection{Comparative Statics: Marginal Cost}

 Figure~\ref{Fig:compStat1} plots the cumulative distribution function
of the mean field equilibrium population state for~$\kappa = 0.05$ and for different values of the
marginal cost~$c$ at the end of the~$1000$-th iteration. Observe that
as the marginal cost increases from~$c = 0.005$ to $c = 0.05$, the
mass of the equilibrium population state shifts to lower security
levels.  This is as predicted by Theorem \ref{th:comp-stat}: at a
higher marginal cost, it is costly to 
maintain a higher security level, and hence players tend to invest less---
resulting in an equilibrium distribution with substantial weight at
lower states.  Note that even small changes in the marginal cost can
significantly shift the equilibrium profile.

\begin{figure}
\centering
\includegraphics[width=5.5in]{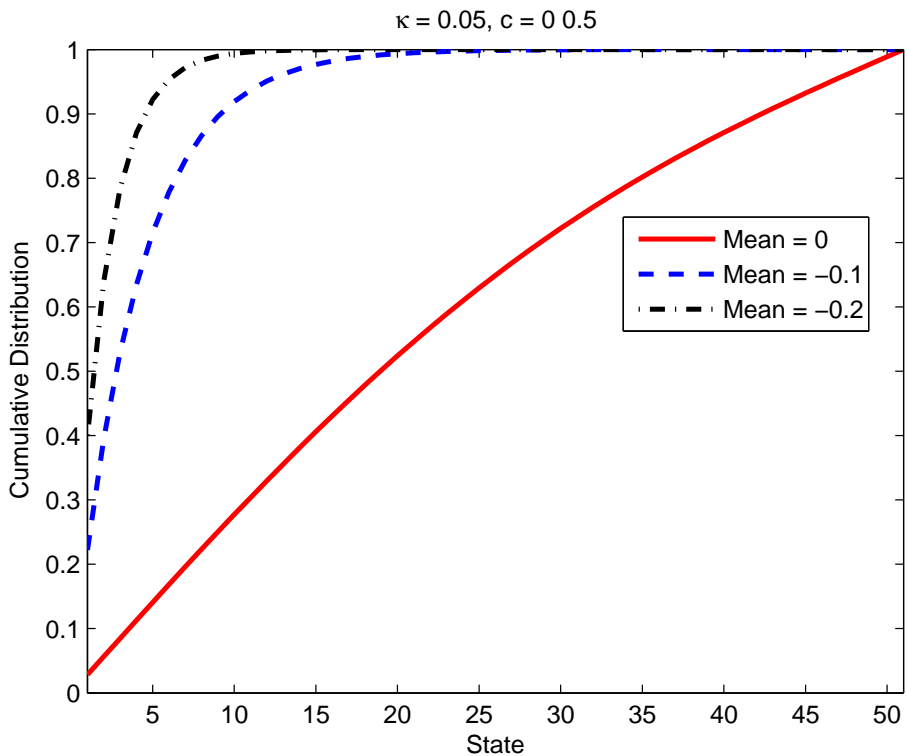}
\caption{Cumulative distribution function of population state at $t =
  1000$ under L-MLD for~$\kappa = 0.05$  and marginal cost~$c = 0.05$,
  with different distributions for $W$.  In all cases $q_0 = 0.2$.  The
  solid line is for $q_{-1} 
  = q_1 = 0.4$; the dashed line is for $q_{-1} = 0.45, q_1 = 0.35$; and
  the dashed dotted line is for $q_{-1} = 0.5, q_1 = 0.3$.}
\label{Fig:CdfVsNoise}
\end{figure}

\subsection{Comparative Statics: Transition Kernel}

Figure~\ref{Fig:CdfVsNoise} plots a different kind of comparative
statics result, where we plot the cumulative distribution function
of the mean field equilibrium population state 
for different noise distributions.  We observe that as the mean of the
noise distribution becomes negative, the equilibrium distribution
tends to concentrate over lower states. One can interpret this
negative mean as the tendency of the player's security to deteriorate over
time, e.g., because the anti-virus software installed on a machine
becomes outdated.  Thus, each player needs to constantly take an action
to maintain its security level. For a fixed marginal cost of action,
a more negative drift results in players moving toward lower
security levels. Note that 
even for small negative drift in the noise distribution, the
cumulative distribution rapidly concentrates over lower states. Thus
in order to maintain a desired security level in a network, a network
administrator should try to ensure that an individual player's
security level does not depreciate quickly over time.

\subsection{Heterogeneity}
\label{subsec:hetero}
We now consider a model where players may be
heterogeneous.  Specifically, we consider two types of players whose payoff
function is parameterized by an {\em interaction}
parameter~$\delta$. This interaction parameter controls the effect of
the security level of other players on the payoff of an individual
player. The payoff function is then given by:
\[
\pi\left(x,a,f; \delta\right) =  -p(x) - \delta (1- p(x))\eta(f) - c a,
\]
where $\eta(f)$ is the mean of $f$.  A higher value of~$\delta$
implies that a player frequently interacts with other players in the
network. Thus, the average security level of other players has a higher
impact on its own security.  As discussed in the conclusion, all the
results of our paper continue to apply for a model with heterogeneous
players.

\begin{figure}
\centering
\includegraphics[width=5.5in]{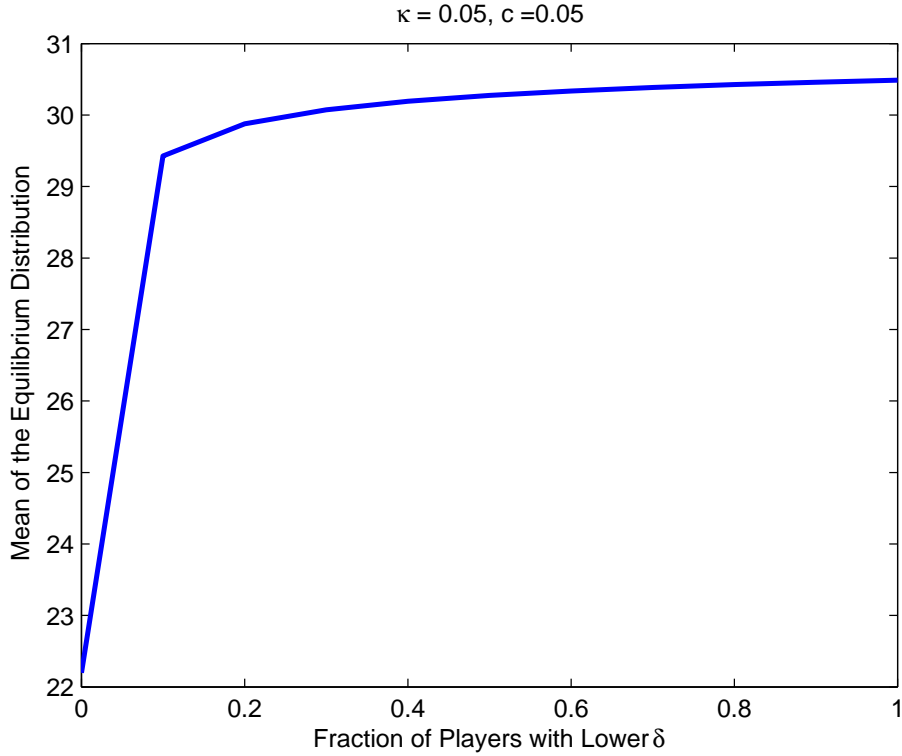}
\caption{Mean of the mean field equilibrium population state vs. the fraction of players
  with low interaction parameter, i.e., $\delta = \delta_L$.  Here
  $\kappa = 0.05$ and $c = 0.05$.}
\label{Fig:hetero}
\end{figure}

For the purposes of this numerical
example, we consider two values of the interaction parameter---a low value
given by~$\delta_{L} = 0.1$ and a high value given by~$\delta_{H} =
0.9$. Figure~\ref{Fig:hetero} plots the mean of the mean field equilibrium population
state as the fraction of players with lower~$\delta$
increases. When all players have a high interaction parameter, they
interact with each other more often. Hence each player has a high
probability of a bad event, and so feels that a personal investment in
security is not likely to be particularly productive.  A ``tragedy of
the commons'' ensues, and the resulting mean population state is quite
low in equilibrium.   On the other hand, even for a small fraction
of players with the low interaction parameter, the mean security level
of {\em all} players
increases.  This is because those players feel a positive benefit to
investing their own security level, and thus encourage others to
invest as well---even those with the high interaction parameter.  Thus
in a real network, slightly limiting interaction 
between players can have a significant impact on the overall security
level of the system. 

\begin{figure}
\centering
\includegraphics[width=5.5in]{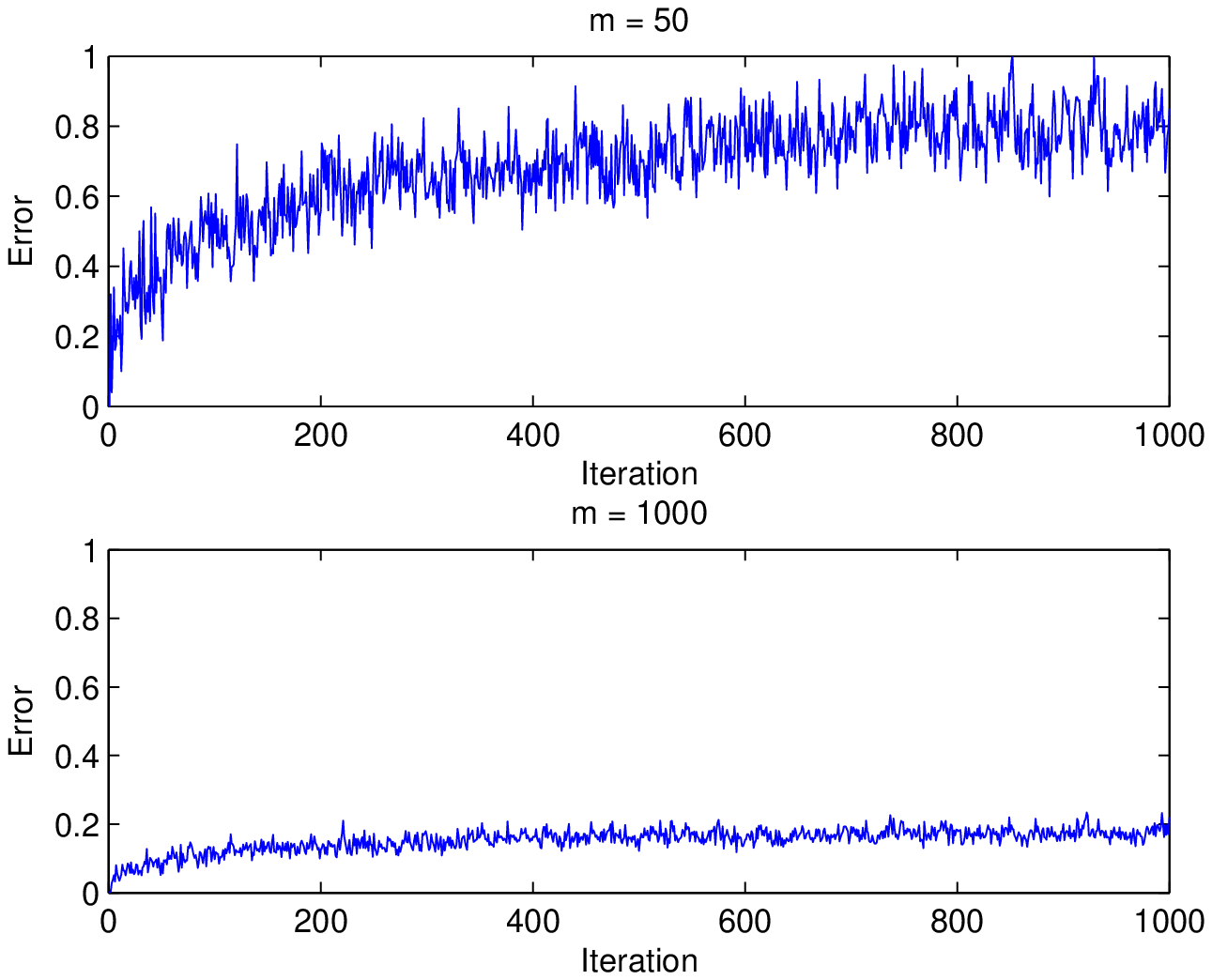}
\caption{Total variation distance between the actual distribution and
  the empirical distribution for L-MLD with~$m=50$ and L-MLD with~$m=1000$ players.  Here
  $\kappa = 0.05$ and $c = 0.05$.}
\label{Fig:errors}
\end{figure}

\subsection{Convergence of L-MLD: Finite vs. Mean Field}

As noted above, the mean field equilibrium is computed
using L-MLD.  In such dynamics, agents compute their
current optimal strategy assuming that the population state remains
fixed for all
future time. Using this computed optimal strategy, the next 
population state is computed using
equation~\eqref{eq:brd-cd}. This process is repeated until
convergence.  In this computation, the next population state is a
{\em deterministic} function of the previous distribution, and it
depends on the optimal strategy via the transition kernel; this is a
consequence of the mean field limit.

Here we ask the question: what happens if finitely many players use
L-MLD?  In other words, each player $i$ in a game with $m$ players
observes the
true population state at time $t$, $f_{-i,t}^{(m)}$; and then executes
L-MLD with respect to this population state. Errors are then
introduced because the next population state 
is {\em stochastic} in a finite system.  (See discussion at the end of
Section \ref{sec:convergence}.)

In
Figure~\ref{Fig:errors}, we plot the total variation error between
successive population states, for $m = 50$ and $m = 1000$ players.
As we observe, for a small number
of players ($m = 50$), the error can be high, and
accumulates as time passes.   However, for a large number of
players ($m = 1000$), this error is considerably reduced.

\begin{figure}
\centering
\includegraphics[width=5.5in]{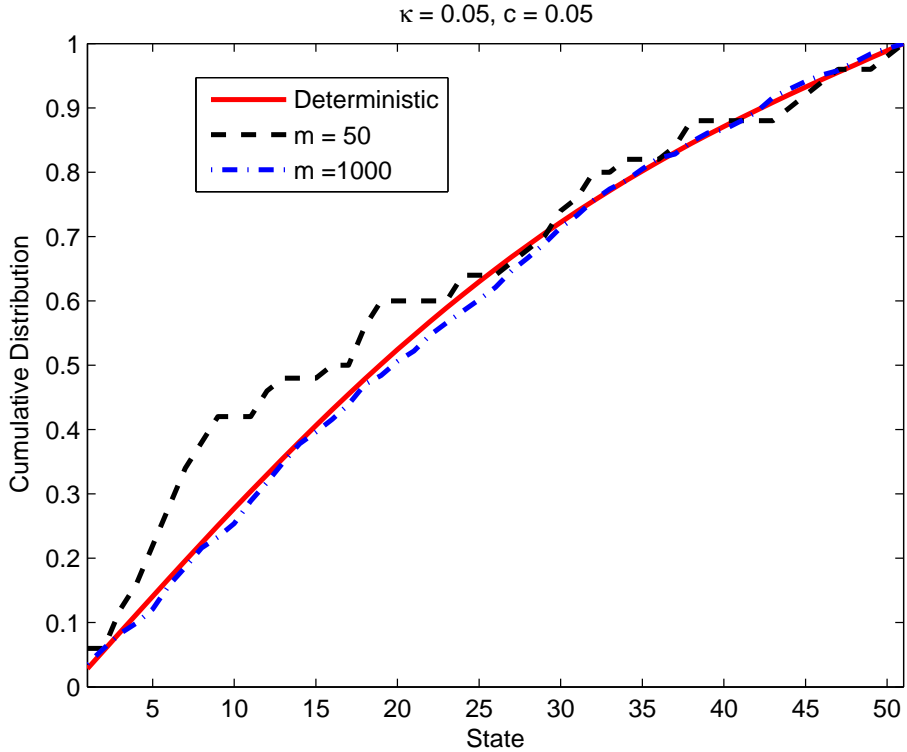}
\caption{Cumulative distribution function of the population state at $t =
  1000$ under L-MLD for~$\kappa = 0.05$  and marginal cost~$c =
  0.05$.  The plot includes the population state obtained under the
  deterministic  mean field L-MLD; L-MLD
with $m = 50$ players; and L-MLD with $m = 1000$ players.}
\label{Fig:stoc-cdf}
\end{figure}

This effect is
also seen in the limiting population state. In
Figure~\ref{Fig:stoc-cdf}, we plot the cumulative distribution
function at~$t = 1000$ for three cases: the mean field L-MLD; L-MLD
with $m = 50$ players; and L-MLD with $m = 1000$ players.
As
we observe, for $m=1000$, the population state is very close to the
population state obtained using the mean field deterministic update
(cf. \eqref{eq:brd-cd}).

% The numerical example presented here highlights the utility of using mean field equilibrium to understand the impact of model primitives on the equilibrium behavior of the system. As seen from Figure~\ref{Fig:stoc-cdf}, the dynamics presented in this paper give good predictions about the actual behavior for moderately sized population. Furthermore, the mean field equilibrium allows us to study the impact of model primitive such as changing marginal cost of action (Figure~\ref{Fig:compStat1}), noise distribution (Figure~\ref{Fig:CdfVsNoise}), heterogeneous players (Figure~\ref{Fig:hetero}), etc. on the over all equilibrium behavior of large number of players. Since the computation of the equilibrium can be done relatively fast, the mean field equilibrium concept is a relatively quick and accurate tool to analyze large population stochastic games.

%% file: conclusion.tex
\section{Conclusion}
\label{sec:conclusion}

This paper has considered existence of mean field equilibrium in games
that exhibit {\em strategic complementarities} in the states of the
players.  Our proofs exploit monotonicity and complementarity
properties of the model primitives to demonstrate that there exist
both a ``largest'' and ``smallest'' mean field equilibrium among all
equilibria where the strategy is nondecreasing in the state.  Further,
we demonstrate that there exist natural myopic learning dynamics that
converge to these equilibria.  Finally, we apply our results in the
context and illustrate how specific examples of games with
complementarities may be analyzed using our techniques.

% We note that the state space, action space, and kernel parameter space are
% all scalar in our model.  The state space can be multidimensional,
% as long as the dynamics of each component of the state are
% decoupled; otherwise, in general, the collection of population
% states may not form a complete lattice \cite{echenique_2003}.  Our
% proof techniques rely heavily on the scalar nature of the action and
% kernel parameter spaces (cf. Lemma \ref{lem:HCprops}); relaxing
% these conditions remains an open direction.

We conclude by noting two extensions that can be developed for the
models described here.

\begin{enumerate}
\item {\em Types}.  In our model players are homogeneous; however,
  this is not a consequential restriction, and is made primarily for
  convenience. In Section~\ref{subsec:hetero}, we considered a numerical
  example with heterogeneous players. 
  More
  generally, we can extend the definition of a stochastic game by
  assuming that there exists a finite {\em type space}
  $\Delta$, with $\pi(x,a,f;\delta)$ and $\mbf{P}(\cdot | x,
  a,f;\delta)$ the payoff and transition kernel, respectively, of a
  type~$\delta$ player.  Further, we assume that the probability a
  player is of type~$\delta$ is given by $\psi(\delta)$.  With this
  extension, as long as the conditions of Definition~\ref{def:basic-cd} are satisfied for each~$\delta$, it is
  straightforward to extend our existence, convergence, and
  comparative statics results.  The main technical issue is that now a
  mean field equilibrium must provide an optimal strategy~$\mu_\delta$
  and population state $f_\delta$ for each~$\delta$.  We omit the
  details.  
\item {\em Multidimensional state and action spaces}.  A more difficult extension
  involves models where the state and action spaces may be
  multidimensional lattices.  The main challenge here arises because the set of
  distributions on a multidimensional compact lattice $\mc{X}$ is {\em not}
  generally a lattice in the first order stochastic dominance
  ordering; see \cite{kamae_1977} for details.  However, first order
  stochastic dominance {\em does} give a closed partial order on the
  set of distributions on $\mc{X}$.

  We can leverage this fact as follows.  \newrj{Suppose that in addition to
  the conditions of Definition \ref{def:basic-cd}, the action set is
  a fixed lattice $\mc{A}$ for all $x$, i.e., $\mc{A}(x) = \mc{A}$ for all
  $x$.}\footnote{\newrj{We employed the total ordering of $\mc{A}(x)$ in proving
    the value function $V^*(x|f)$ is nondecreasing in $x$, via Lemma
    \ref{lem:DPincrdiff-cd}; however, if $\mc{A}$ does not depend on $x$, then it is
    straightforward to check that $V^*(x|f)$ is nondecreasing in $x$ even if
    $\mc{A}$ is multidimensional.}}  Further, suppose the model
  primitives (payoff and transition kernel) are all continuous in
  state, action, and population state---i.e., Assumption
  \ref{as:continuity-cd} is satisfied.  Then Kleene's fixed point
  theorem \citep{kleene_1971} can be used to establish existence of, and
  convergence to, a mean field equilibrium.  Kleene's fixed point
  theorem states that if $X$ is a space with a closed partial order
  and a smallest element, then any monotone continuous function from $X$
  to itself possesses a fixed point.  We omit the details of
  this argument, as it is essentially identical to our preceding
  development.  

  We do emphasize, however, that our analysis of {\em separable} stochastic
  games is intimately tied to the assumption that state and action
  spaces are single-dimensional.  In particular, our proof techniques
  rely heavily on the scalar nature of the action and kernel parameter
  spaces (cf. Lemma \ref{lem:HCprops}); relaxing these conditions
  remains an open direction.
\end{enumerate}
%\item The state space, action space, and kernel parameter space are
%  all scalar in our model.  The state space can be multidimensional,
%  as long as the dynamics of each component of the state are
%  decoupled; otherwise, in general, the collection of population
%  states may not form a complete lattice \cite{echenique_2003}.  Our
%  proof techniques rely heavily on the scalar nature of the action and
%  kernel parameter spaces (cf. Lemma \ref{lem:HCprops}); relaxing
%  these conditions remains an open direction.
%\item Under a slight strengthening of the continuity assumption
%  (Assumption \ref{as:continuity}), 
%  it can be shown that the {\em asymptotic Markov equilibrium}
%  property holds: as the number of players grows large, mean field
%  equilibrium becomes a more and more accurate approximation to Markov
%  perfect equilibrium.  (In a Markov perfect equilibrium, agents
%  strategies' are optimal given both their own state as well as the
%  actual state evolution of other agents.)  For further details on
%  general conditions under which the AME property holds, the reader is
%  referred to \cite{adlakha_2010}.
%\end{enumerate}

%% file: appendix.tex
\begin{APPENDICES}

\section{Proofs: Section \ref{sec:existthmproof}}

We start with the lemma that demonstrates that optimal strategies exist, 
and can be identified via Bellman's equation; the proof uses standard results
from dynamic programming.

\begin{lemma}
\label{lem:bellman-cd}
For each $f \in \mfr{F}$, $\mc{P}(f)$ is nonempty. Furthermore, $\mu \in \mc{P}(f)$ if and only if  for each $x$:
\begin{align*}
\mu(x) \in \arg\max_{a \in \mc{A}(x)}\left\{\pi(x, a, f) + \beta \int_{\mc{X}} V^*(x' | f) \P(dx' | x, a, f) \right\}
\end{align*}
\end{lemma}
\proof{Proof.}
Throughout the proof we employ Definition \ref{def:basic-cd}.  In
particular, observe that the payoff is continuous on a compact set $\mathsf{A}$ and thus
for fixed~$f$, the payoff $\pi(x, a, f)$ is bounded. In addition, for each
fixed $x$, the next
state is drawn from a countable set by assumption.  Thus consider
maximization of the expected discounted profit over {\em all} possible
(randomized, history-dependent) strategies; by standard results in the
theory of dynamic programming (see \citealp{Bertsekas78}), it
can be shown that if there exists an optimal strategy, there must
exist an optimal stationary, nonrandomized, Markov strategy---i.e., in
our terminology, an
oblivious strategy.  Further, $V^*$  satisfies Bellman's equation:
\begin{align}
\label{eq:bellman-cd}
V^*(x | f) = \sup_{a \in \mc{A}(x)} \left\{ \pi(x,a,f) + \beta
  \int_{\mc{X}} V^*(x' | f) \P(dx' | x, a, f)\right\};
\end{align}
and an oblivious strategy is optimal if and only if it
attains the maximum on the right hand side of the preceding
expression for every $x$. 

Observe also that for fixed $f$, $V^*(\cdot | f)$ is bounded, since the per stage
payoffs are bounded and the discount factor is less than one. Since the
transition probability $\P(\cdot | x, a, f)$ is continuous w.r.t. the
topology of weak convergence, we conclude the objective function in
\eqref{eq:bellman-cd} is continuous in $a$. Because $\mc{A}(x)$ is compact,  the
maximum is achieved for every $x$, and thus at least one optimal
strategy must exist---i.e., $\mc{P}(f)$ is nonempty. This proves the lemma. \hfill \halmos
\endproof

The next three lemmas combine to show that the value function $V^{*}(x | f)$ 
has increasing differences in~$x$ and $f$.

\begin{lemma}
\label{lem:Tincrdiff-cd}
Suppose that $U(x | f)$ is a nondecreasing bounded function in $x$ and has increasing differences in $x$ and $f$. Define
\begin{align}
\label{eqn:T}
T(x, a, f) = \int_{\mc{X}}U(x'|f)\P(dx'| x, a, f).
\end{align}
Then $T(x, a, f)$ is nondecreasing in $x$ \newrj{and $a$}, supermodular in $(x, a)$ and has increasing differences in $(x, a)$ and $f$.
\end{lemma}
\proof{Proof.}
By Definition~\ref{def:basic-cd}, $\P\left(\cdot | x, a ,f \right)$
is stochastically nondecreasing in $x$ \newrj{and $a$} and stochastically supermodular in
$(x, a)$. Since $U(x|f)$ is a nondecreasing bounded function, it
follows from the definition of stochastically nondecreasing and
stochastically supermodular that $T(x, a, f)$ is nondecreasing in~$x$
\newrj{and $a$} and supermodular in $(x, a)$ for fixed $f$.

Fix $\hat{x} \geq x$, $\hat{a} \geq a$, and $\hat{f} \succeq_{\SD} f$ and define
\begin{align*}
\hat{T}(x, a, f, g) = \int_{\mc{X}}U(x|f)\P(dx| x, a, g).
\end{align*}
To prove $T(x, a, f)$ has increasing differences in $(x, a)$ and $f$, it suffices to show that
\begin{align}
\label{eqn:Tincrdiff-cd-to-prove}
\hat{T}(\hat{x}, \hat{a}, \hat{f}, \hat{f}) - \hat{T}(x, a, \hat{f}, \hat{f}) \geq \hat{T}(\hat{x}, \hat{a},f, f) - \hat{T}(x, a, f, f).
\end{align}
Let us fix $g$; since $U(x|f)$ has increasing differences in $x$ and $f$, $U(x|\hat{f}) - U(x|f)$ is a nondecreasing function of $x$. Since $\P(\cdot| \hat{x}, a, g) \succeq_{\SD} \P(\cdot| x, a, g)$ by Definition~\ref{def:basic-cd}, it follows that:
\begin{align}
\label{eqn:Tincrdiff-cd-temp1}
\hat{T}(\hat{x}, a, \hat{f}, g) - \hat{T}(\hat{x}, a, f, g) \geq \hat{T}(x, a, \hat{f},g) - \hat{T}(x,a, f, g).
\end{align}
Also by Definition~\ref{def:basic-cd}, $\P(\cdot| \hat{x}, \hat{a}, g) \succeq_{\SD} \P(\cdot| \hat{x}, a, g)$ which implies that
\begin{align}
\label{eqn:Tincrdiff-cd-temp2}
\hat{T}(\hat{x}, \hat{a}, \hat{f}, g) - \hat{T}(\hat{x}, \hat{a}, f, g) \geq \hat{T}(\hat{x}, a, \hat{f},g) - \hat{T}(\hat{x},a, f, g).
\end{align}
Using equations~\eqref{eqn:Tincrdiff-cd-temp1} and \eqref{eqn:Tincrdiff-cd-temp2} and rearranging the terms, we get that
\begin{align}
\label{eqn:Tincrdiff-cd-temp3}
\hat{T}(\hat{x}, \hat{a}, \hat{f}, g) - \hat{T}(x, a, \hat{f},g) \geq \hat{T}(\hat{x}, \hat{a}, f, g) - \hat{T}(x,a, f, g).
\end{align}
Now let $\hat{g}\succeq_{\SD} g$ and note that $\P(\cdot|x, a, g)$
has increasing differences in $(x, a)$ and $g$ by
Definition~\ref{def:basic-cd}. Also, note that $U(x|f)$ is a bounded nondecreasing function of $x$. This implies that $\hat{T}(x, a, f, g)$ has increasing differences in $(x, a)$ and $g$. That is,
\begin{align}
\label{eqn:Tincrdiff-cd-temp4}
\hat{T}(\hat{x}, \hat{a}, \hat{f}, \hat{g}) - \hat{T}(x, a, \hat{f},\hat{g}) \geq \hat{T}(\hat{x}, \hat{a}, \hat{f}, g) - \hat{T}(x,a, \hat{f}, g).
\end{align}
From equations~\eqref{eqn:Tincrdiff-cd-temp3} and \eqref{eqn:Tincrdiff-cd-temp4}, we get that for any $\hat{x} \geq x$, $\hat{a} \geq a$, $\hat{f} \succeq_{\SD} f$, and $\hat{g} \succeq_{\SD} g$ we have
\begin{align*}
\hat{T}(\hat{x}, \hat{a}, \hat{f}, \hat{g}) - \hat{T}(x, a, \hat{f},\hat{g}) \geq \hat{T}(\hat{x}, \hat{a}, f, g) - \hat{T}(x,a, f, g).
\end{align*}
Taking $\hat{f} = \hat{g}$ and $f = g$ in the above equation shows that equation~\eqref{eqn:Tincrdiff-cd-to-prove} is true which proves the lemma. \hfill \halmos
\endproof

\begin{lemma}
\label{lem:DPincrdiff-cd}
Suppose that $U(x|f)$ is a nondecreasing bounded function in $x$ that has increasing differences in $x$ and $f$. Define
\begin{align*}
U^{*}(x|f) = \sup_{a \in \mc{A}(x)}\left\{ \pi(x, a, f) + \beta \int_{\mc{X}}U(x'|f)\P\left(dx'|x, a, f\right)\right\}.
\end{align*}
Then $U^{*}(x|f)$ is nondecreasing in $x$ and has increasing differences in $x$ and $f$.
\end{lemma}
\proof{Proof.}
Define 
\begin{align*}
 W(x, a, f) &= \pi(x, a, f) + \beta \int_{\mc{X}}U(x'|f)\P\left(dx'|x, a, f\right)\\
 & = \pi(x, a, f) + \beta T(x, a, f).
\end{align*}
From Lemma~\ref{lem:Tincrdiff-cd}, we know that $T(x, a, f)$ is
nondecreasing in $x$ \newrj{and $a$}, supermodular in $(x, a)$ and has increasing
differences in $(x, a)$ and $f$. From Definition~\ref{def:basic-cd},
we get that $W(x, a, f)$ is nondecreasing in $x$, supermodular in $(x,
a)$ and has increasing differences in $(x, a)$ and
$f$. 

%\delrj{Since $W$ is nondecreasing in $x$ for all $a$ and $f$, it follows that
%$\sup_{a \in \mc{A}(x)} W(x,a,f)$ is also nondecreasing in $x$ for all
%$f$.}

\newrj{
We now show that $U^*(x|f)$ is nondecreasing in $x$ for fixed $f$.  Fix $x' \geq
x$, and choose $\hat{a} \in \mc{A}(x)$ such that $W(x,\hat{a},f) =
\sup_{a \in \mc{A}(x)} W(x,a,f)$; such an action exists since $\pi$
and $\P$ are continuous in $a$ and $\mc{A}(x)$ is compact.  
Similarly, fix
$\hat{a}'$ such that $\pi(x,\hat{a}',f) =
\sup_{a \in \mc{A}(x')} \pi(x,a,f)$.

We consider two cases.  If
$\hat{a} \in \mc{A}(x')$, then:
\begin{align*}
U^*(x|f)& = \pi(x, \hat{a}, f) + \beta T(x, \hat{a}, f) \\
&\leq \pi(x', \hat{a}, f) + \beta T(x', \hat{a}, f)\\
&\leq \sup_{a \in \mc{A}(x')} \pi(x', a, f) + \beta T(x', a, f) = U^*(x'|f),
\end{align*}
where the first inequality follows since $\pi$ and $T$ are both
nondecreasing in $x$.

On the other hand, if $\hat{a} \not\in \mc{A}(x')$, then it follows
that $\hat{a}' > \hat{a}$, since $\mc{A}$ is a nondecreasing
correspondence.  (Note that this step uses the fact that the action
set is totally ordered.)  Thus:
\begin{align*}
U^*(x|f)& = \pi(x, \hat{a}, f) + \beta T(x, \hat{a}, f) \\
&\leq \pi(x', \hat{a}', f) + \beta T(x, \hat{a}, f)\\
&\leq \pi(x', \hat{a}', f) + \beta T(x', \hat{a}', f)\\
&\leq \sup_{a \in \mc{A}(x')} \pi(x', a, f) + \beta T(x', a, f) = U^*(x'|f),
\end{align*}
Here the first inequality follows since $\sup_{a \in \mc{A}(x)}
\pi(x,a,f)$ is nondecreasing in $x$, and by our choice of $\hat{a}'$;
and the second inequality follows since $T$ is nondecreasing in $x$
and $a$.  We conclude that $U^*(x | f)$ is nondecreasing in $x$ for
fixed $f$.
}

To prove that $U^{*}(x|f)$ has increasing differences, we reason using an argument similar to Lemma $A.1$ of~\cite{Hopenhayn92}. Let $x_{2} \geq x_{1}$ and $f_{2} \succeq_{\SD} f_{1}$. To prove that $U^{*}(x|f)$ has increasing differences, we need to show that
\begin{align}
\label{eqn:DPincrdiff-cd-to-prove}
U^{*}(x_{2}|f_{2}) - U^{*}(x_{1}|f_{2}) \geq U^{*}(x_{2}|f_{1}) - U^{*}(x_{1}|f_{1}).
\end{align}

To economize on notation, given two actions $a,a'$, we let $a \vee a'
= \sup\{a,a'\}$, and let $a \wedge a' = \inf\{a,a'\}$.  Fix $a_{1} \in
\mc{A}(x_1)$ and $a_{2} \in \mc{A}(x_2)$.  Since $\mc{A}(x)$ is a nondecreasing correspondence, we
have $a_{1} \vee a_{2} \in \mc{A}(x_2)$ and $a_{1} \wedge a_{2} \in
\mc{A}(x_1)$. Thus we have: 
\begin{align*}
U^{*}(x_{1}|f_{1}) + U^{*}(x_{2}|f_{2}) &\geq W(x_{1}, a_{1}\wedge a_{2}, f_{1}) + W(x_{2}, a_{1}\vee a_{2}, f_{2})\\
& = W(x_{1}, a_{1}\wedge a_{2}, f_{1}) + W(x_{2}, a_{1}\vee a_{2}, f_{2}) + W(x_{1}, a_{1}\wedge a_{2}, f_{2}) \\
& \quad \quad \quad - W(x_{1}, a_{1}\wedge a_{2}, f_{2}) \\
 &= W(x_{1}, a_{1}\wedge a_{2}, f_{1}) + W(x_{1} \vee x_{2}, a_{1}\vee a_{2}, f_{2})  + W(x_{1} \wedge x_{2}, a_{1}\wedge a_{2}, f_{2}) \\
 & \quad \quad   - W(x_{1}, a_{1}\wedge a_{2}, f_{2}) 
\end{align*}
The last equality follows from that fact that $x_{1} \vee x_{2} = x_{2}$ and $x_{1}\wedge x_{2} = x_{1}$. Since $W(x, a, f)$ is supermodular in $(x,a)$ we get that 
\begin{align*}
U^{*}(x_{1}|f_{1}) + U^{*}(x_{2}|f_{2}) &\geq W(x_{1}, a_{1}\wedge a_{2}, f_{1}) + W(x_{2}, a_{2}, f_{2}) + W(x_{1}, a_{1}, f_{2}) \\
& \quad \quad \quad - W(x_{1}, a_{1}\wedge a_{2}, f_{2}), \\
& = W(x_{1}, a_{1}, f_{2}) + W(x_{2}, a_{2}, f_{1}) -  W(x_{2}, a_{2}, f_{1}) + W(x_{2}, a_{2}, f_{2}) \\
&\quad \quad \quad + W(x_{1}, a_{1}\wedge a_{2}, f_{1})- W(x_{1}, a_{1}\wedge a_{2}, f_{2})\\
%& = W(x_{1}, a_{1}, f_{2}) + W(x_{2}, a_{2}, f_{1}) + W(x_{2}, a_{2}, f_{2}) - W(x_{1}, a_{1}\wedge a_{2}, f_{2}) \\
%&\quad \quad \quad  - \left(W(x_{2}, a_{2}, f_{1}) - W(x_{1}, a_{1}\wedge a_{2}, f_{1})\right)\\
& \geq W(x_{1}, a_{1}, f_{2}) + W(x_{2}, a_{2}, f_{1}).
\end{align*}
Here the last inequality follows from the fact that $W(x, a, f)$ has increasing differences in $(x, a)$ and $f$. Taking the supremum over $a_{1}$ and $a_{2}$ in the above inequality we get 
\begin{align*}
U^{*}(x_{1}|f_{1}) + U^{*}(x_{2}|f_{2}) &\geq U^{*}(x_{1} | f_{2}) + U^{*}(x_{2} | f_{1}) 
\end{align*}
which implies equation~\eqref{eqn:DPincrdiff-cd-to-prove}, and thus $U^{*}(x|f)$ has increasing differences in $x$ and $f$. This proves the lemma. \hfill \halmos
\endproof

\begin{lemma}
\label{lem:V*incrdiff-cd}
$V^{*}(x|f)$ is nondecreasing in $x$ and has increasing differences in $x$ and $f$.
\end{lemma}
\proof{Proof.}
Let $V_0(x | f ) = 0$ for all $x$, and let:
\begin{align}
V_{k+1}(x | f) = \sup_{a \in \mc{A}(x)} \left\{ \pi(x,a,f) + \beta
  \int_{\mc{X}} V_k(x' | f) \P(dx' | x, a, f) \right\};
\end{align}
this is value iteration.  By the preceding lemma,
every $V_k$ is nondecreasing in $x$ and has increasing differences in $x$ and $f$. Under our assumptions, value
iteration converges  
starting from the zero function \citep{bertsekas_2001_2}, i.e., for all $x$,
$V_k(x | f) \to V^*(x|f)$ as $k \to \infty$.  Since monotonicity and increasing differences are
preserved upon taking limits, we conclude
$V^*(x | f)$ is nondecreasing in $x$ and has increasing differences in $x$ and $f$. \hfill \halmos
\endproof

We now apply Topkis' Theorem in the next lemma to conclude
the set of optimal strategies is monotone.

\begin{lemma}
\label{lem:Gammaincr-cd}
For each $x$ and $f$, define the set $\Omega(x, f)$ as:
\begin{align}
\label{eq:Gamma-cd}
\Omega(x, f) = \arg \max_{a \in \mc{A}(x)} \left\{ \pi(x,a, f) + \beta \int_{\mc{X}} V^*(x' | f) \P(dx' |x, a, f) \right\}.
\end{align}
Then $\Omega$ is nondecreasing in $(x,f)$.

Further:
\[ \ol{p}(f)(x) = \sup \Omega(x,f);\ \  \ul{p}(f)(x) = \inf
\Omega(x,f), \]
where $\ol{p}$ and $\ul{p}$ are the strategies defined in
\eqref{eq:ulpolp-cd}.  Both $\ol{p}(f)$ are $\ul{p}(f)$ are
nondecreasing in $f$, and for fixed~$f$ both strategies are also
nondecreasing in $x$. 
\end{lemma}

\proof{Proof.}
Observe that $\pi(x, a, f)$ is supermodular in $(x, a)$ and has
increasing differences in $(x,a)$ and~$f$ (by
Definition~\ref{def:basic-cd}); and $T(x, a, f)$ is supermodular in
$(x,a)$ and has increasing differences in $(x, a)$ and $f$, where $T$
is defined in equation~\eqref{eqn:T}, with $U = V^{*}$ (by
Lemma~\ref{lem:Tincrdiff-cd} and~\ref{lem:V*incrdiff-cd}).  Further,
$\mc{A}(x)$ is an increasing correspondence.  By Topkis' Theorem (Theorem 2.8.1 in \citealp{topkis_1998}), we conclude that $\Omega(x,f)$ is nondecreasing in $(x,f)$ wherever it is nonempty.
By Lemma \ref{lem:bellman-cd}, however, the maximum on the right hand
side in \eqref{eq:Gamma-cd} is always achieved, so $\Omega(x,f)$ is nondecreasing everywhere.

To conclude the proof, observe that by Lemma \ref{lem:bellman-cd},
$\ol{p}(f)$ must be the strategy that takes the largest action in
$\Omega(x,f)$ for each $x$, and $\ul{p}(f)$ must be the strategy that
takes the smallest action in $\Omega(x,f)$ for each $x$.  Monotonicity
of $\ol{p}$ and $\ul{p}$ follows from the monotonicity properties of $\Omega$. \hfill \halmos
\endproof

We now turn our attention to $\mc{D}$. Given any strategy $\mu$ and
population state $f$, define a map $Q_{\mu, f}: \mfr{F} \rightarrow
\mfr{F}$ according to the
kernel induced by $\mu$ and $f$, i.e., for all Borel sets S:
\[ 
Q_{\mu, f}(g)(S) = \int_{\mc{X}}\P(S|x, \mu(x), f)\; g(dx).
\]
(This is equation \eqref{eq:Qgamma-cd}.)

\begin{lemma}
\label{lem:Qgamma-cd}
Suppose $f' \succeq_{\SD} f$, $g' \succeq_{\SD} g$, and $\mu'\unrhd \mu$, and both $\mu'$ and $\mu$ are nondecreasing, then $Q_{\mu', f'}(g') \succeq_{\SD} Q_{\mu, f}(g)$.
\end{lemma}
\proof{Proof.}
Let $\phi$ be a bounded nondecreasing real-valued function on
$\mc{X}$.   We need to show that for every $x \in \mc{X}$, we have
\begin{align}
\label{eqn:Qgamma-cd-to-prove}
\int_{\mc{X}} \int_{\mc{X}} \phi(y) \P\left(dy|x, \mu'(x),
f'\right)g'(dx) \geq \int_{\mc{X}}\int_{\mc{X}}\phi(y) \P\left(dy|x, \mu(x), f\right)g(dx).
\end{align}
Let us define 
\begin{align*}
H(x; \mu, f) = \int_{\mc{X}}\phi(y)\; \P(dy|x, \mu(x), f).
\end{align*}
Observe that:
\[  \int_{\mc{X}}\int_{\mc{X}} \phi(y) \P(dy|x, \mu(x),
f)\; g(dx) = \int_{\mc{X}} H(x; \mu,f)\; g(dx).\]
Let $x' \geq x$ and note that $\mu$ is a nondecreasing function of $x$. From Definition~\ref{def:basic-cd}, we know that $\P(\cdot|x, a, f)$ is stochastically nondecreasing in $(x, a)$, which implies that $\P(\cdot |x', \mu(x'), f) \succeq_{\SD} \P(\cdot |x, \mu(x), f)$. Since $\phi$ is a nondecreasing function, we get that $H(x'; \mu, f) \geq H(x; \mu, f)$. 

From Definition~\ref{def:basic-cd}, we know that $\P(\cdot|x, a, f)$ is nondecreasing in $a$ and $f$. Thus, for any fixed $x$, we have
\begin{align*}
\P(\cdot|x, \mu'(x), f') \succeq_{\SD} \P(\cdot|x, \mu(x), f).
\end{align*}
This along with the fact that $\phi$ is a nondecreasing function implies that $H(x;\mu', f') \geq H(x; \mu, f)$ for every fixed $x\in \mc{X}$. 

We now reason as follows:
\begin{align*}
\int_{\mc{X}}H(x; \mu', f')g'(dx) &\geq \int_{\mc{X}}H(x; \mu, f)g'(dx)\\
&\geq \int_{\mc{X}}H(x; \mu, f)g(dx).
\end{align*}
Here the first inequality follows from the fact that $H(x;\mu', f')
\geq H(x; \mu, f)$ and the second inequality follows from that fact
that $H(x'; \mu, f) \geq H(x; \mu, f)$ for $x' \geq x$, and that $g' \succeq_{\SD} g$. This proves equation~\eqref{eqn:Qgamma-cd-to-prove} and hence proves the lemma.\hfill \halmos
\endproof

\begin{lemma}
\label{lem:Dgamma-cd}
Fix $\mu \in \mfr{M}_{O}$ and $f \in \mfr{F}$, and suppose $\mu$ is nondecreasing in
$x$.  Then $\mc{D}(\mu, f)$ is a nonempty complete lattice.
Further, $\ul{d}(\mu, f)$ and
$\ol{d}(\mu, f)$ (as defined in \eqref{eq:uldold-cd}) exist and are both
invariant distributions 
of the Markov 
process induced by $\mu$ and $f$ (cf. \eqref{eqn:inv-dist}).

Finally, if $f' \succeq_{\SD} f$ and 
$\mu' \unrhd \mu$ and both $\mu$ and $\mu'$ are
nondecreasing, then $\ul{d}(\mu', f') \succeq_{\SD} \ul{d}(\mu, f)$ and
$\ol{d}(\mu', f') \succeq_{\SD} \ol{d}(\mu, f)$.
\end{lemma}

\proof{Proof.}
By the preceding lemma, $Q_{\mu, f}(g)$ is nondecreasing in $g$.  By Tarski's theorem, the set of fixed points of 
$Q_{\mu, f}$ is a nonempty complete
lattice.  View the set of nondecreasing strategies
  $\mu$ as a partially ordered set $\mfr{M}_{O}$, with the
  coordinate-wise ordering~$\unrhd$.  Then $Q_{\mu, f}(\cdot)$ is a
  nondecreasing function on $\mfr{M}_{O} \times \mfr{F} \times \mfr{F}$, by the
  preceding lemma.  Theorem 2.5.2 in~\cite{topkis_1998} generalizes
  Tarski's theorem to fixed points of a nondecreasing function
  parametrized by a partially ordered set ($\mfr{M}_{O} \times \mfr{F}$); one consequence of this
  generalization is that the largest and smallest fixed points are
  nondecreasing in the parameter.  This generalization directly
  implies that both $\ul{d}(\mu, f)$ and $\ol{d}(\mu, f)$ are
  nondecreasing in $\mu$ and $f$. \hfill \halmos
\endproof

In the next lemma we establish existence of fixed points of $\Phi$,
thus proving Theorem \ref{th:existence-cd}.

\begin{lemma}
\label{lem:fixedpts-cd}
Let $\ul{\Phi}(f)$ and
$\ol{\Phi}(f)$ be defined as in \eqref{eq:ulphiolphi-cd}.  Then
$\ul{\Phi}(f), \ol{\Phi}(f) \in \Phi(f)$. Further, both are nondecreasing in
$f$, and thus the sets of their fixed 
points are each nonempty complete lattices.

Thus there exists a mean field equilibrium for the stochastic game
with complementarities
$\Gamma$: in particular, if $f$ is a 
fixed point of $\ul{\Phi}$ (resp., $\ol{\Phi}$), then $(\ul{p}(f), f)$
(resp., $(\ol{p}(f), f)$) is a mean field equilibrium.  
\end{lemma}

\proof{Proof.}
That $\Phi(f)$ is nonempty follows by Lemmas \ref{lem:bellman-cd} and
\ref{lem:Dgamma-cd}.  Observe that if $f' \succeq_{\SD} f$, then
$\ul{p}(f') \unrhd \ul{p}(f)$ by Lemma \ref{lem:Gammaincr-cd}.
Further, $\ul{p}(f')$ and $\ul{p}(f)$ are both nondecreasing in $x$ as
well, so by Lemma~\ref{lem:Dgamma-cd},  $\ul{d}(\ul{p}(f'), f')
\succeq_{\SD} \ul{d}(\ul{p}(f), f)$, establishing that $\ul{\Phi}$ is
monotone.  That $\ul{\Phi}(f) \in \Phi(f)$ follows from the
definition.  The proof for $\ol{\Phi}(f)$ is identical.  The
conclusion regarding fixed points follows from Tarski's theorem. 
\hfill \halmos
\endproof

\section{Proofs: Section \ref{ssec:ordering-cd}}

\proof{Proof of Corollary \ref{cor:ordmfe-cd}.}
Since $\ol{p}(f) = \sup \mc{P}(f)$, and $\mu \in
\mc{P}(f)$, we must have $\ol{p}(f) \unrhd \mu$.  Similarly
$\mu \unrhd \ul{p}(f)$.  Now since $\ol{p}(f)$ and $\ul{p}(f)$ are
both nondecreasing strategies, by Lemma \ref{lem:Dgamma-cd} we conclude
that $ \ol{\Phi}(f) = \ol{d}(\ol{p}(f), f) \succeq_{\SD} \ol{d}(\mu, f) \succeq_{\SD} f$,
where the last inequality follows because $\ol{d}(\mu, f) = \sup
\mc{D}(\mu, f)$, and $f \in \mc{D}(\mu, f)$.  A similar
argument shows that $f \succeq_{\SD} \ul{\Phi}(f)$.  The result
now follows from Theorem 2.5.1 in \cite{topkis_1998}, which is a sharper
version of Tarski's fixed point theorem; in particular, that statement
shows that $\ol{f} = \sup \{ f' : \ol{\Phi}(f) \succeq_{\SD} f \}$,
and $\ul{f} = \inf \{ f' : f \succeq_{\SD} \ul{\Phi}(f) \}$.  Since
$f$ is contained in both the former and the latter sets, we conclude
$\ul{f} \preceq_{\SD} f \preceq_{\SD} \ol{f}$.  The result regarding
strategies now follows from monotonicity of $\ul{p}$ and $\ol{p}$, and
the fact that $\ul{p}(f) \unlhd \mu \ol{p}(f)$. \hfill \halmos
\endproof

\section{Proofs: Section \ref{sec:convergence}}

We start with two essential lemmas.

\begin{lemma}
\label{lem:f_mu_seqs}
Suppose that $f_0 \preceq_{\SD} f_1
\preceq_{\SD} f_2 \cdots$, and $\mu_0 \unlhd \mu_1 \unlhd
\mu_2 \cdots$.  Then there exists a distribution $f^*$ and a strategy
$\mu^*$ such that $f_t$ converges
weakly to $f^*$ as $t 
\to \infty$, and $\mu_t$ converges pointwise to $\mu^*$ as $t \to
\infty$.
\end{lemma}

\proof{Proof.}
Note that since the set $\mfr{M}_{O}$ is compact and $\mu_t$ is a nondecreasing sequence
of policies, there must exist a pointwise limit $\mu^*$.  
Next, consider the distribution functions $F_t$ corresponding to
the measures $f_t$.  By Prohorov's theorem, there exists a measure
$f^* \in \mfr{F}$ and a subsequence $t_k$ such that $F_{t_k}(x) \to
F^*(x)$ at all points of continuity of $F^*$, where $F^*$ is the
distribution function corresponding to $f^*$.  Since $f_t$ is a monotone sequence, we have
$F_t(x) \geq F_{t+1}(x)$ for all $x$, so in fact $F_t(x) \to F^*(x)$
at all points of continuity of $F^*$.  Thus $f_t$ converges weakly to
$f^*$, as required.\hfill \halmos
\endproof

\begin{lemma}
\label{lem:continuity}
Suppose Assumption \ref{as:continuity-cd} holds.  Then $\ul{p}(f)$ and
$\ol{p}(f)$ (cf. \eqref{eq:ulpolp-cd}) are both continuous in $f$, and
$Q_{\mu, g}(f)$ (cf. \eqref{eq:Qgamma-cd}) is continuous in $\mu$,
$g$, and $f$, where we endow $\mfr{F}$ with 
the topology of weak convergence, and 
$\mfr{M}_{O}$ with the topology of pointwise convergence.
\end{lemma}

\proof{Proof.}
Under
Assumption \ref{as:continuity-cd}, the first result follows using Theorem
1 of \cite{Dutta94}, which establishes upper semicontinuity of
$\Omega(x,f)$ in $f$ (where $\Omega$ is defined as in
\eqref{eq:Gamma-cd}).  From this it follows that $\ul{p}(f)$ and 
$\ol{p}(f)$ are continuous in $f$.

Next we show that $Q_{\mu, g}(f)$ as defined in
\eqref{eq:Qgamma-cd} is continuous in $\mu$, $g$ and $f$, 
where we endow $\mfr{F}$ with the topology of weak convergence, and
$\mfr{M}_{O}$ with the topology of pointwise convergence.  
Suppose that $g_{n} \to g$ (weakly), $f_n \to f$
(weakly), and that $\mu_n \to \mu$ (pointwise).  Fix a bounded
function $\phi$ on $\mc{X}$, and define 
\[
H(x; \mu, g) = \int_{\mc{X}}\phi(y)\mbf{P}(dy  |  x, \mu(x), g).
\]
For every $x$, $H(x; \mu_{n}, g_{n}) \to H(x; \mu, g)$ by continuity of 
$\mbf{P}(\cdot  |  x, a, g)$ in $a$ and $g$. Thus by Theorem~5.5 of
\cite{Billingsley68}, we have:
\[
\int_{\mc{X}}H(x; \mu_{n}, g_{n})f_{n}(dx) \to \int_{\mc{X}}H(x; \mu, g)f(dx).
\]
The left hand side is the expected value of $\phi$ under
$Q_{\mu_n, g_{n}}(f_n)$, and the right hand side is the expected value
of $\phi$ under $Q_{\mu, g}(f)$, so (weak) continuity of $Q_{\mu,
  g}(f)$ is proved.\hfill \halmos
\endproof

\proof{Proof of Proposition \ref{prop:conv-l-brd}.}
Since $f_{t+1} = \ul{\Phi}(f_t)$, and $\ul{\Phi}$ is monotone by Lemma
\ref{lem:fixedpts-cd}, it follows that $f_0 \preceq_{\SD} f_1
\preceq_{\SD} f_2 \cdots$. Since $\mu_t = \ul{p}(f_t)$, and $\ul{p}$
is monotone in $f_t$ by Lemma \ref{lem:Gammaincr-cd}, it follows that $\mu_0 \unlhd \mu_1 \unlhd
\mu_2 \cdots$.  Finally, since $\ul{p}(f)(x)$ is nondecreasing in $x$
for every $f$, it follows that $\mu_t$ is nondecreasing.  By Lemma
\ref{lem:f_mu_seqs}, there exists a limit $(\mu^*, f^*)$. Since
every $\mu_t$ is nondecreasing in $x$, the limit $\mu^*$ must be
nondecreasing in $x$ as well.

We now show that if Assumption \ref{as:continuity-cd} holds, then the
limit point $(\mu^*, f^*)$ is the smallest mean field equilibrium.  By
Lemma \ref{lem:continuity}, both $\ul{p}(f)$ and $Q_{\mu,g}(f)$ are
continuous.  This implies that $\mu_t = \ul{p}(f_t) \to \ul{p}(f^*)$ as $t \to
\infty$, so $\mu^* = \ul{p}(f^*)$.  Further, since $f_{t+1} = \ul{d}(\mu_t, f_t)$, it follows
that $Q_{\mu_t, f_t}(f_{t+1}) = f_{t+1}$.  Taking limits on the left
and right, we have $Q_{\mu^*, f^*}(f^*) = f^*$, i.e., $f^* \in
\mc{D}(\mu^*, f^*)$.  Thus we conclude $(\mu^*, f^*)$ is a mean field
equilibrium.

Let $\ul{f}$ be the smallest fixed point of $\ul{\Phi}(f)$, as defined
in \eqref{eq:ulfolf-cd}.  Observe that at time 0,
$f_0 \preceq_{\SD} \ul{f}$, since $f_0$ is the smallest distribution in the
lattice $\mfr{F}$.  Since $\Phi$ is monotone, $f_t \preceq_{\SD} \ul{f}$ for
all $t$.  Since $f_t$ converges weakly to $f^*$, we conclude $f^*
\preceq_{\SD} \ul{f}$.  On the other hand, observe that $\mu^*$ is
nondecreasing, so by Corollary \ref{cor:ordmfe-cd}, we have $\ul{f}
\preceq_{\SD} f^*$---i.e., $f^* = \ul{f}$, as required.  
\hfill \halmos
\endproof

\proof{Proof of Proposition \ref{prop:conv-cd}.}
We proceed by induction.  First note that by Lemma \ref{lem:Gammaincr-cd},
$\ul{p}(f)(x)$ is a nondecreasing strategy in $x$ for each $f$, so every $\mu_t$ is
nondecreasing.  We start by observing that $f_0$ is the smallest
distribution in $\mfr{F}$ in the $\succeq_{\SD}$ ordering, so $f_1
\succeq_{\SD} f_0$ trivially.  Since $\ul{p}$ is monotone in $f$ by
Lemma \ref{lem:Gammaincr-cd} we
have $\mu_0 = \ul{p}(f_0) \unlhd \ul{p}(f_1) = \mu_1$.  

So now suppose that $f_0 \preceq_{\SD} f_1
\preceq_{\SD}  \cdots \preceq_{\SD} f_t$, and $\mu_0 \unlhd \mu_1 \unlhd
\cdots \unlhd \mu_t$.  Define $Q_{\mu_t, f_{t}}$ according to~\eqref{eq:Qgamma-cd}.  Then by Lemma \ref{lem:Qgamma-cd},
$Q_{\mu_t, f_{t}}$ is nondecreasing; since $f_{t+1} = Q_{\mu_t, f_{t}}(f_t)$,
we conclude $f_{t+1} \succeq_{\SD} f_t$.  The same argument as the
preceding paragraph then yields $\mu_{t+1} \unrhd \mu_t$, as
required.  Applying Lemma \ref{lem:f_mu_seqs} yields the convergence
result; note that $\mu^*$ must be nondecreasing, since every $\mu_t$
is nondecreasing.

From Lemma \ref{lem:continuity}, if Assumption \ref{as:continuity-cd}
holds, we conclude that $\mu^* =
\ul{p}(f^*)$, and $Q_{\mu^*, f^{*}}(f^*) = f^*$ --- i.e., $f^* \in
\mc{D}(\mu^*, f^{*})$.  Thus $(\mu^*, f^*)$ is a mean field equilibrium.

Let $\ul{f}$ be the smallest fixed point of $\ul{\Phi}(f)$, as defined
in \eqref{eq:ulfolf-cd}.  Observe that at time 0,
$f_0 \preceq_{\SD} \ul{f}$, since $f_0$ is the smallest distribution in the
lattice $\mfr{F}$.  Thus $\mu_0 = \ul{p}(f_0) \unlhd
\ul{p}(\ul{f})$, so $f_1 = Q_{\mu_0, f_{0}}(f_0) \preceq_{\SD}
Q_{\ul{p}(\ul{f}), \ul{f}}(\ul{f}) = \ul{f}$, where the last equality follows
since $\ul{f}$ must be an invariant distribution associated with $\ul{p}(\ul{f})$ and $\ul{f}$.  
Proceeding inductively, we have $f_t \preceq_{\SD} \ul{f}$ for all
$t$.  Since $f_t$ converges weakly to $f^*$, we conclude $f^*
\preceq_{\SD} \ul{f}$.  On the other hand, observe that $\mu^*$ is
nondecreasing, so by Corollary \ref{cor:ordmfe-cd}, we have $\ul{f}
\preceq_{\SD} f^*$---i.e., $f^* = \ul{f}$, as required.  
\hfill \halmos
\endproof

\section{Proof: Section \ref{sec:comp-stat}}

\proof{Proof Sketch for Theorem \ref{th:comp-stat}.}
Let $\mc{P}(f;\theta)$ denote the set of optimal oblivious strategies
for an agent given population state $f$ in the game $\Gamma(\theta)$;
and let $\mc{D}(\mu,f;\theta)$ denote the set of invariant
distributions induced by the strategy $\mu$ and population state $f$
in the game $\Gamma(\theta)$.  Let $\ul{p}(f;\theta)$ and
$\ol{p}(f;\theta)$ be defined as in 
\eqref{eq:ulpolp-cd} using $\mc{P}(f;\theta)$; and similarly, let $\ul{d}(\mu, f;\theta)$ and
$\ol{d}(\mu, f;\theta)$ be defined as in \eqref{eq:uldold-cd} using
$\mc{D}(\mu, f;\theta)$.  Using exactly the same reasoning as in
the proof of Theorem \ref{th:existence-cd}, under Assumption
\ref{as:comp-stat}, it follows that both $\ul{p}(f;\theta)$ and
$\ol{p}(f;\theta)$ are
nondecreasing in $f$ and $\theta$, and nondecreasing in state; and
further, that $\ul{d}(\mu, f; \theta)$ and $\ol{d}(\mu, f;\theta)$ are
nondecreasing in $\mu$, $f$, and $\theta$, when restricted to
strategies $\mu$ that are nondecreasing in the state.  Letting
$\ul{\phi}(f;\theta) = \ul{d}(\ul{p}(f;\theta), f;\theta)$ and
$\ol{\phi}(f;\theta) = \ol{d}(\ol{p}(f;\theta), f;\theta)$, it follows
that $\ul{\phi}$ and $\ol{\phi}$ are nondecreasing in $f$ and
$\theta$.  By Theorem 2.5.2 in~\cite{topkis_1998}, the largest and
smallest fixed points of $\ul{\phi}(f;\theta)$ and
$\ol{\phi})(f;\theta)$ are nondecreasing in~$\theta$.  The result
follows.
\hfill \halmos
\endproof

\section{Proofs: Section \ref{sec:actions}}

\proof{Proof Sketch for Theorem \ref{th:existence-action}.}  Analogous to the proof of Theorem
\ref{th:existence-cd}, we define $\ul{p}(\alpha) = \inf
\mc{P}(\alpha)$, and $\ol{p}(\alpha) = \sup \mc{P}(\alpha)$ (with the
$\inf$ and $\sup$ taken coordinatewise); and $\ol{d}(\mu, \alpha) =
\sup \mc{D}(\mu, \alpha)$, and $\ul{d}(\mu, \alpha) = \inf \mc{D}(\mu,
\alpha)$.  Lemmas \ref{lem:bellman-cd}, \ref{lem:Tincrdiff-cd},
\ref{lem:DPincrdiff-cd}, \ref{lem:V*incrdiff-cd}, and
\ref{lem:Gammaincr-cd} remain identical, except that the population
state $f$ is replaced by the population action distribution $\alpha$.
Next, we define $Q_{\mu, \alpha}(g)$ as:
\[ Q_{\mu, \alpha}(g)(S) = \int_{\mc{X}} \P(S | x, \mu(x), \alpha)\;
g(dx). \]
An identical arguments to the proof of Lemma \ref{lem:Qgamma-cd} then
shows that if $\alpha' \succeq_{\SD} \alpha$, $g' \succeq_{\SD} g$, and $\mu'\unrhd \mu$, and both $\mu'$
and $\mu$ are nondecreasing, then $Q_{\mu', f'}(g') \succeq_{\SD}
Q_{\mu, f}(g)$.  (Here $\alpha'$ and $\alpha$ are population action
distributions; $g'$ and $g$ are population states; and $\mu'$ and $\mu$
are oblivious strategies.)  It follows that Lemma \ref{lem:Dgamma-cd} remains identical
 as well, with the population state $f$ replaced by $\alpha$.

Finally, we turn our attention to $\hat{\mc{D}}$.  Suppose that $\mu'
\unrhd \mu$ (where $\mu$ and $\mu'$ are both measurable oblivious strategies),
and $f' \succeq_{\SD} f$ (where $f$ and $f$' are both population
states).   Suppose $\phi : \textsf{A} \to \reals$ is nondecreasing.
Let $\hat{\alpha} =  \hat{\mc{D}}(\mu, f)$, and let $\hat{\alpha}' =
\hat{\mc{D}}(\mu', f')$.
Then:
\begin{align*}
\int_{\textsf{A}} \phi(a)\; \alpha'(da) &= \int_{\mc{X}} \phi(\mu'(x))\;
f'(dx) \\
&\geq \int_{\mc{X}} \phi(\mu(x))\;
f'(dx) \\
&\geq \int_{\mc{X}} \phi(\mu(x))\;
f(dx) = \int_{\textsf{A}} \phi(a)\; \alpha(da).
\end{align*}
The first inequality follows because $\mu'(x) \geq \mu(x)$ for all
$x$, and $\phi$ is nondecreasing.  The second inequality follows
because $\mu(x)$ is nondecreasing, so $\phi(\mu(x))$ is nondecreasing
in $x$; and $f' \succeq_{\SD} f$.  It follows from this argument that
$\hat{\mc{D}}(\mu', f') \succeq_{\SD} \hat{\mc{D}}(\mu, f)$,
establishing that $\hat{\mc{D}}$ is monotone as well.

So now we define two functions $\ul{\Phi} : \mfr{F}_{\mathsf{A}} \to
\mfr{F}_{\mathsf{A}}$ and $\ol{\Phi}  : \mfr{F}_{\mathsf{A}} \to \mfr{F}_{\mathsf{A}}$, analogous
to the definitions in \eqref{eq:ulphiolphi-cd}.  Let
$\ul{\Phi}(\alpha) = \hat{\mc{D}}(\ul{p}(\alpha),
\ul{d}(\ul{p}(\alpha), \alpha))$; and let $\ol{\Phi}(\alpha) = \hat{\mc{D}}(\ol{p}(\alpha),
\ol{d}(\ol{p}(\alpha), \alpha))$.  Then both $\ul{\Phi}$ and
$\ol{\Phi}$ are monotone maps on the nonempty complete lattice
$\mfr{F}_{\mathsf{A}}$, so the set of fixed points of both maps are
nonempty complete lattices by Tarski's fixed point theorem.  In particular, any one of these fixed
points is a mean field equilibrium of the given action-coupled
stochastic game.\hfill \halmos
\endproof

\section{Proofs: Section \ref{sec:existence}}

\proof{Proof of Lemma \ref{lem:HCprops}.}
Since $h$ is concave in $a$ for fixed $x$ (Definition \ref{def:basic}),
it is also continuous in $a$ for fixed $x$; since $\mc{A}$ is a
compact interval, it follows by the intermediate value theorem that
$H(x)$ is a compact interval.  Now 
if $x' \geq x$, then $h(x',a) \geq h(x,a)$ for all $a$ (by Definition
\ref{def:basic}), so we conclude $H(x)$ is nondecreasing in $x$.  

Since $h(x,a)$ is nondecreasing in both $x$ and $a$, and $c(a)$ is
nondecreasing in $a$, it follows that $C(x,\hat{h})$ is nondecreasing in $\hat{h}$ when
$x$ is fixed, and nonincreasing in $x$ when $\hat{h}$ is fixed.
\newrj{Equation~\eqref{eq:Cmin} also follows since $c(a)$ is nondecreasing in
$a$.}  Convexity in $\hat{h}$
follows by standard results in convex optimization:
since we restrict attention to $\hat{h} \in H(x)$ and $h$ and $c$ are
both nondecreasing in $a$, we can rewrite the constraint $h(x,a) =
\hat{h}$ as $h(x,a) \geq \hat{h}$ in the definition of
$C(x,\hat{h})$, i.e., for $\hat{h} \in H(x)$ we have:
\[ C(x,\hat{h}) = \inf_{a \in \mc{A} : h(x,a) \geq \hat{h}} c(a). \] 
Now since $C(x,\hat{h})$ is defined via minimization of a convex objective
function over a convex feasible region parametrized by $\hat{h}$, it
is convex in $\hat{h}$ \citep{Bertsekas09}.

Finally, we establish the claim of decreasing differences.  Fix
$x,x',\hat{h}$, and $\hat{h}'$ as in the statement of the lemma.
Define $\alpha_1, \alpha_2, \alpha_3$, and $\alpha_4$ as optimizing values
of $a$ in the definition of $C(x,\hat{h})$, $C(x',\hat{h})$,
$C(x,\hat{h}')$, and $C(x',\hat{h}')$, respectively.  We have $h(x,\alpha_1) =
h(x', \alpha_2) = \hat{h}$, and $h(x,\alpha_3) = h(x', \alpha_4) =
\hat{h}'$.  

Observe that since $\hat{h}' > \hat{h}$ and $h$ is nondecreasing in
action, $\alpha_4 > \alpha_2$,
and $\alpha_3 > \alpha_1$.  Further, since $h$ is nondecreasing in
$x$, we have $\alpha_4 \leq \alpha_3$ and $\alpha_2 \leq \alpha_1$.
Let $\delta = \alpha_4 - \alpha_2$.  Define $g(a) = -h(x, -a)$ for $a
\in -\mc{A}$; then observe that $g$ is a convex, nondecreasing
function on $-\mc{A}$. By Lemma \ref{lem:convex} (see below), we have:
\[ g(-\alpha_2) - g(-\alpha_4) \geq g(-\alpha_1) -
g(-\alpha_1-\delta). \]
In terms of $h$, this implies:
\begin{equation}
\label{eq:hineq}
h(x, \alpha_4) - h(x,\alpha_2) \geq h(x, \alpha_1 + \delta) -
h(x,\alpha_1). 
\end{equation}
We can now show that $\alpha_4 - \alpha_2
\leq \alpha_3 - \alpha_1$.   We have:
\begin{align*}
\hat{h}' - \hat{h} &= h(x', \alpha_4) - h(x',\alpha_2)\\
&\geq h(x,\alpha_4) - h(x,\alpha_2)\\
&\geq h(x,\alpha_1 + \delta) - h(x,\alpha_1).
\end{align*}
Here the first inequality follows by supermodularity of $h$ in $(x,a)$ (Definition
\ref{def:basic}), and the second inequality follows by
\eqref{eq:hineq}.  Since $h(x,\alpha_1) = \hat{h}$
and $h(x,\alpha_3) = \hat{h}'$, and $h$ is nondecreasing in action, it
follows that $\alpha_1 + \delta 
\leq \alpha_3$, i.e., $\alpha_3 - \alpha_1 \geq \alpha_4 - \alpha_2$.  

The result now follows by another application of Lemma
\ref{lem:convex} (see below), which implies:
\[ c(\alpha_3) - c(\alpha_1) \geq c(\alpha_4) - c(\alpha_2), \]
or equivalently,
\[ C(x, \hat{h}') - C(x, \hat{h}) \geq C(x', \hat{h'}) - C(x',
  \hat{h}),\]
as required. \hfill \halmos
\endproof

\begin{lemma}
\label{lem:convex}
Let $S \subset \reals$ be convex, and suppose $g: S \to \reals$ is a nondecreasing
convex function.  Fix $x,x',y,y' \in S$, such that $y \geq x$, $y' > y$,
$x' > x$, and $y' - y \geq x' - x$.  Then:
\[ g(y') - g(y) \geq g(x') - g(x). \]
\end{lemma}

\proof{Proof.}
Define $\hat{y}' = y + (x' - x)$.  Clearly $y < \hat{y}' \leq y'$, so
$\hat{y}' \in S$.  
Observe that $x < y$ implies $x' < \hat{y}'$, so we can choose
$\alpha, \delta \in (0,1)$ such that: 
\begin{align*}
\alpha x + (1 - \alpha)\hat{y}' &= x';\\
\delta x + (1 - \delta)\hat{y}' &= y.
\end{align*}
In particular, it follows that $\alpha = (\hat{y}'-x')/(\hat{y}'-x)$, and $\delta =
(\hat{y}'-y)/(\hat{y}'-x)$.  Since $\hat{y}' - x' = y - x$, we conclude $\alpha +
\delta = 1$.  Applying convexity we have:
\begin{align*}
 g(x') &\leq \alpha g(x) + (1 - \alpha)g(\hat{y}'); \\
 g(y) &\leq \delta g(x) + (1 - \delta)g(\hat{y}')
\end{align*}
Adding these together, and using the fact that $\alpha + \delta = 1$,
we conclude $g(x') + g(y) \leq g(x) + g(\hat{y}')$, or:
\[ g(x') - g(x) \leq g(\hat{y}') - g(y). \]
Finally, since $g$ is nondecreasing, $g(\hat{y}') \leq g(y')$, and the
result follows.\hfill \halmos
\endproof

\proof{Proof of Proposition \ref{prop:transform}.}
We simply check the conditions outlined in Definition
\ref{def:basic-cd}.  Observe that $v(x,f)$ has is nondecreasing in $x$
and has increasing differences
in $x$ and $f$ by assumption, and further, $\sup_{x \in \mc{X}}
|v(x,f)| < \infty$.  In addition $C(x, \hat{h})$ is convex in
$\hat{h}$ and nonincreasing in $x$, and has decreasing
differences in $x$ and $\hat{h}$ by Lemma \ref{lem:HCprops}.  It
follows that $\hat{\pi}(x, \hat{h}, f)$ is nondecreasing in $x$;
continuous in~$\hat{h}$; supermodular in $(x,
\hat{h})$ (the latter as it is separable in $x$ and
$\hat{h}$); and has increasing differences in $(x, \hat{h})$
and~$f$. Furthermore, for fixed $\hat{h}$ and~$f$, $\sup_{x \in
  \mc{X}}\pi(x, \hat{h}, f) < \infty$.  Thus the first two conditions
of Definition \ref{def:basic-cd} are satisfied.

Next, we consider the transition kernel.  Here the desired properties
follow by assumption: $\hat{\P}(\cdot  | x, \hat{h}, f)$ is
trivially stochastically supermodular in $(x, \hat{h})$, since it does
not depend on $x$ and $\hat{h}$ is scalar.  By assumption the kernel
is stochastically nondecreasing in $\hat{h}$ and $f$,  continuous
in $\hat{h}$, and has increasing differences in $\hat{h}$ and $f$.
%Countable support is also part of Definition \ref{def:basic}.

Finally, note $H(x)$ is a compact interval, $H$ is
nondecreasing, and $H(x) \subset [\ul{h}, \ol{h}]$ for all $x$, which is also
compact.  \newrj{Further, observe that for all $x$ and $f$:
\[ \sup_{\hat{h} \in H(x)} \hat{\pi}(x, \hat{h}, f) = v(x,f) -
\inf_{\hat{h} \in H(x)} C(x, \hat{h}) = v(x,f) - c(\ul{a}),\]
where the last step follows by Lemma \ref{lem:HCprops}.  Since $v$ is
nondecreasing in $x$, it follows that $\sup_{\hat{h} \in H(x)}\pi(x, \hat{h}, f)$ is nondecreasing in $x$.}
It follows that $\Gamma$ is a stochastic game with
complementarities, as required.\hfill \halmos
\endproof

\proof{Proof of Theorem \ref{th:separable}.}
Let $\hat{\Gamma}$ be the equivalent stochastic game with
complementarities constructed in Proposition \ref{prop:transform}.
Suppose that $(\hat{\mu}, \hat{f})$ is a mean field equilibrium of
$\hat{\Gamma}$; note that in this case $\hat{\mu}$ is a strategy where
$\hat{\mu}(x) \in H(x)$ for all $x \in \mc{X}$.  

Define a new strategy $\mu : \mc{X} \to \mc{A}$ as follows.  For each
$x$, let $\mu(x)$ be an action such that $h(x, \mu(x)) =
\hat{\mu}(x)$.  That is, we choose the action $\mu(x)$ to yield
exactly the kernel parameter $\hat{\mu}(x)$.  Then observe that
$\pi(x, \mu(x), g) = \hat{\pi}(x, \hat{\mu}(x), g)$, for all $x$ and
$g$.  Since $\hat{\mu}$ is an optimal oblivious strategy for a player
given population state $f$
in $\hat{\Gamma}$, by construction of $\hat{\Gamma}$ the strategy $\mu$
maximizes the expected discounted payoff to a player given $f$ in the
original game $\Gamma$.  Further, since 
$\P(\cdot | x, \mu(x), g) = \mbf{\hat{P}}( \cdot | x,
\hat{\mu}(x), g)$, it follows that $f$ is an invariant distribution of
the strategy~$\mu$.  Thus $(\mu, f)$ is a mean field equilibrium of the
game $\Gamma$, as required.\hfill \halmos
\endproof

\begin{lemma}
\label{lem:Tincrdiff}
Suppose that $\Gamma$ is a stochastic game satisfying all the conditions in Definition \ref{def:basic}, {\em
  except} that $v$ is not necessarily nondecreasing in $x$.  Suppose
in addition that $\P(\cdot | \hat{h},f)$ does not depend on $f$.
Then $\int_{\mc{X}} V^*(x' | f) \P(dx' | \hat{h})$ has increasing
differences in $\hat{h}$ and $f$.
\end{lemma}

\proof{Proof.}
Let $U(x | f)$ be any function that has increasing differences in $x$
and $f$. Define:
\[ T(\hat{h}, f) = \int_{\mc{X}} U(x' | f) \P(dx' | \hat{h}). \]
Fix $\hat{h}' \geq \hat{h}$ and $f' \succeq_{\SD} f$.  Then since $U(x|f)$
has increasing differences in $x$ and $f$, $U(x | f') - U(x|f)$ is a
nondecreasing function of $x$.  Since $\P(\cdot | \hat{h}')
\succeq_{\SD} \P(\cdot | \hat{h})$ by Definition \ref{def:basic}, it follows that:
\begin{align*}
 \int_{\mc{X}} (U(x|f') - U(x|f)) \P(dx|\hat{h}') \geq 
 \int_{\mc{X}} (U(x|f') - U(x|f)) \P(dx|\hat{h}).
\end{align*}
This is exactly the relationship that $T(\hat{h}', f') - T(\hat{h}',f)
\geq T(\hat{h},f') - T(\hat{h},f)$, so $T$ has increasing differences
in $\hat{h}$ and $f$.

The remainder of the proof follows Lemma~\ref{lem:DPincrdiff-cd}, and~\ref{lem:V*incrdiff-cd}.  First, the same approach as the proof of Lemma \ref{lem:DPincrdiff-cd}
can be used to show that $U^*(x | f)$ has increasing differences in
$x$ and $f$, where:
\[ U^{*}(x|f) = \sup_{\hat{h} \in H(x)}\left\{ v(x, f) - C(x, \hat{h})
+ \beta \int_{\mc{X}}U(x'|f)\P\left(dx'|x, \hat{h}\right)\right\}.\]
Finally, value iteration yields that $V^*(x | f)$ has increasing
differences in $x$ and $f$, as is shown in Lemma~\ref{lem:V*incrdiff-cd}.
\hfill \halmos
\endproof

\end{APPENDICES}